\documentclass[journal,twocolumn,10pt]{IEEEtran}
\usepackage{graphicx}
\usepackage{amssymb}
\usepackage{mathtools}
\usepackage{dsfont}
\usepackage{cite}
\usepackage{stfloats}
\usepackage{subfigure}
\usepackage{psfrag}
\usepackage[mathscr]{euscript}
\usepackage{acronym}  
\usepackage{amsmath}
\usepackage{algorithm}
\usepackage[noend]{algpseudocode}
\usepackage{booktabs}
\usepackage{bbm}
\usepackage{verbatim} 
\usepackage{soul}

\makeatletter
\renewcommand{\maketag@@@}[1]{\hbox{\m@th\normalsize\normalfont#1}}%
\makeatother

\makeatletter
\def\BState{\State\hskip-\ALG@thistlm}
\makeatother

\usepackage{float}
\usepackage{balance}

\acrodef{CCDF}{complementary cumulative distribution function}
\acrodef{CF}{characteristic function}
\acrodef{PPP}{Poisson point processe}
\acrodef{RV}{random variable}
\acrodef{i.i.d.}{independent and identically distributed}
\acrodef{PDF}{probability distribution function}
\acrodef{CDF}{cumulative distribution function}
\acrodef{ch.f.}{characteristic function}
\acrodef{AWGN}{additive white Gaussian noise}
\acrodef{SNR}{signal-to-noise ratio}
\acrodef{LRT}{likelihood ratio test}
\acrodef{DRT}{distance ratio test}
\acrodef{GLRT}{generalized likelihood ratio test}
\acrodef{CRLB}{Cram\'{e}r-Rao lower bound}
\acrodef{CRB}{Cram\'{e}r-Rao bound}
\acrodef{ZZLB}{Ziv-Zakai lower bound}
\acrodef{ZZB}{Ziv-Zakai bound}
\acrodef{LOS}{line-of-sight}
\acrodef{ToF}{time-of-flight}
\acrodef{NLOS}{non-line-of-sight}
\acrodef{GDOP}{geometric dilution of precision}
\acrodef{GPS}{Global Positioning System}
\acrodef{FIM}{Fisher information matrix}
\acrodef{PEB}{position error bound}
\acrodef{SPEB}{squared position error bound}
\acrodef{TOA}{time-of-arrival}
\acrodef{TOF}{time-of-flight}
\acrodef{WSN}{wireless sensor network}
\acrodef{MAC}{medium access control}
\acrodef{RSS}{received signal strength}
\acrodef{WAF}{wall attenuation factor}
\acrodef{TDOA}{time difference-of-arrival}
\acrodef{RF}{radiofrequency}
\acrodef{RTT}{round-trip time}
\acrodef{AOA}{angle-of-arrival}
\acrodef{MF}{matched filter}
\acrodef{ED}{energy detector}
\acrodef{ML}{maximum likelihood}
\acrodef{MSE}{mean-square error}
\acrodef{RMSE}{root-mean-square error}
\acrodef{LEO}{localization error outage}
\acrodef{ppm}{part-per-million}
\acrodef{ACK}{acknowledge}
\acrodef{UWB}{Ultrawide bandwidth}
\acrodef{TNR}{threshold-to-noise ratio}
\acrodef{LS}{least squares}
\acrodef{IR-UWB}{impulse radio UWB}
\acrodef{FCC}{Federal Communications Commission}
\acrodef{TH}{time-hopping}
\acrodef{PPM}{pulse position modulation}
\acrodef{MUI}{multi-user interference}
\acrodef{PDP}{power delay profile}
\acrodef{BPZF}{band-pass zonal filter}
\acrodef{SIR}{signal-to-interference ratio}
\acrodef{SINR}{signal-to-interference-plus-noise ratio}
\acrodef{RFID}{radio frequency identification}
\acrodef{WPAN}{wireless personal area network}
\acrodef{WWB}{Weiss-Weinstein bound}
\acrodef{DP}{direct path}
\acrodef{MF}{matched filter}
\acrodef{MMSE}{minimum-mean-square-error}
\acrodef{SBS}{serial backward search}
\acrodef{SBSMC}{serial backward search for multiple clusters}
\acrodef{NBI}{narrowband interference}
\acrodef{WBI}{wideband interference}
\acrodef{INR}{interference-to-noise ratio}
\acrodef{CR}{channel response}
\acrodef{CIR}{channel impulse response}
\acrodef{CR}{channel  response}
\acrodef{RADAR}{radar}
\acrodef{MUR}{Multistatic radar}
\acrodef{JBSF}{jump back and search forward}
\acrodef{HDSA}{high-definition situation-aware}
\acrodef{RRC}{root raised cosine}
\acrodef{ST}{simple thresholding}
\acrodef{BTB}{Bellini-Tartara bound}
\acrodef{P-Max}{$P$-Max}  
\acrodef{MIMO}{multiple-input multiple-output}
\acrodef{MAP}{maximum a posteriori}
\acrodef{FG}{factor graph}
\acrodef{OP}{outage probability}
\acrodef{WED}{wall extra delay}
\acrodef{RMS}{root mean square}
\acrodef{SPAWN}{sum-product algorithm over a wireless network}
\acrodef{MDD}{minimum distance distribution}
\acrodef{MAP}{maximum a posteriori probability}
\acrodef{SAP}{small cell access point}
\acrodef{UE}{user equipment}
\acrodef{MBS}{macro cell base station}
\acrodef{UER}{\ac{UE} Relay}
\acrodef{D2D}{device-to-device}
\acrodef{MBS}{macro base station}
\acrodef{CSI}{channel state information}
\acrodef{OGR}{outage guard region}
\acrodef{FUR}{feasible UER region}
\acrodef{EHR}{energy harvesting region}
\acrodef{EH}{energy harvesting}
\acrodef{D2D-EHSN}{D2D communication provided \ac{EH} small cell network}
\acrodef{D2D-EHHN}{D2D communication provided \ac{EH} heterogeneous network}
\acrodef{3GPP}{3rd Generation Partnership Project}
\acrodef{BS}{base station}
\acrodef{DF}{decode and forward}
\acrodef{CCDF}{complementary cumulative distribution function}
\acrodef{ZF}{zero forcing}
\acrodef{RZF}{regularized zero forcing}
\acrodef{WLLN}{weak law of large number}
\acrodef{SLLN}{strong law of large numbers}
\acrodef{TDD}{Time-division duplex}
\acrodef{EE}{energy efficiency} 
\acrodef{HetNet}{heterogeneous network} 
\acrodef{SCP}{Single Cell Processing}
\acrodef{CBF}{Coordinated Beamforming}
\usepackage{color}
\usepackage{dsfont}
\usepackage{bbm}

\DeclareMathAlphabet{\mathsf}{OML}{cmbr}{m}{it}

\newtheorem{theorem}{\bf Theorem}
\newtheorem{lemma}{\bf Lemma}
\newtheorem{corollary}{\bf Corollary}

\newtheorem{remark}{\bf Remark}

\newcommand{\bd}{\begin{description}}
\newcommand{\ed}{\end{description}}
\newcommand{\be}{\begin{enumerate}}
\newcommand{\ee}{\end{enumerate}}
\newcommand{\bi}{\begin{itemize}}
\newcommand{\ei}{\end{itemize}}
\newcommand{\bl}{\begin{list}}
\newcommand{\el}{\end{list}}
\newcommand{\bt}{\begin{tabbing}}
\newcommand{\et}{\end{tabbing}}

\newcommand{\paperTitle}{ Age of Information Under Frame Slotted ALOHA-Based Status Updating Protocol}

\begin{document}

\title{\paperTitle}



\author{

    Zhiling~Yue,~\IEEEmembership{Student Member, IEEE},
    Howard~H.~Yang,~\IEEEmembership{Member,~IEEE,}
    Meng~Zhang,~\IEEEmembership{Member,~IEEE,}
    and~Nikolaos~Pappas,~\IEEEmembership{Senior~Member,~IEEE}

\thanks{     
    The work of Z. Yue, H. H. Yang, and M. Zhang has been supported by the National Natural Science Foundation of China under Grants 62201504 and 62202427. 
     The work of N. Pappas has been supported in part by the Swedish Research Council (VR), ELLIIT, Zenith, and the European Union (ETHER, 101096526).  (\emph{Corresponding Author: Howard H. Yang.})

    Z. Yue, H. H. Yang and M. Zhang are with the Zhejiang University/University of Illinois at Urbana-Champaign Institute, Zhejiang University, Haining 314400, China (email: chi-ling@zju.edu.cn, haoyang@intl.zju.edu.cn, mengzhang@intl.zju.edu.cn).

    N. Pappas is with the Department of Computer and Information Science, Linköping University, Linköping 58183, Sweden (e-mail: nikolaos.pappas@liu.se).}
}

\maketitle
\acresetall
\thispagestyle{empty}


\begin{abstract}
We propose a frame slotted ALOHA (FSA)-based protocol for 
a random access network where sources transmit status updates to their intended destinations.
We evaluate the effect of such a protocol on the network's timeliness performance using the Age of Information (AoI) metric. 
Specifically, we leverage tools from stochastic geometry to model the spatial positions of the source-destination pairs and capture the entanglement amongst the nodes' spatial-temporal attributes through the interference they caused to each other. 
We derive analytical expressions for the average and variance of AoI over a typical transmission link in Poisson bipolar and cellular networks, respectively.
Our analysis shows that in densely deployed networks, the FSA-based status updating protocol can significantly decrease the average AoI and in addition, stabilizes the age performance by substantially reducing the variance of AoI. 
Furthermore, under the same updating frequency, converting a slotted ALOHA protocol into an FSA-based one always leads to a reduction in the average AoI.
Moreover, implementing FSA in conjunction with power control can further benefit the AoI performance, although the particular values of framesize and power control factor must be adequately tuned to achieve the optimal gain. 
\end{abstract}

\acresetall

\section{Introduction}\label{sec:intro}
Ultra-reliable and low-latency communication (URLLC)  \cite{2018Cross,2021URLLC} is an important technology originated from the intersection of the Internet of Things (IoT) and the tactile Internet, promoting a broad range of real-time applications, such as healthcare \cite{2022health}, autonomous vehicles \cite{autoAoI}, remote sensing and control \cite{2020Control}.
These applications have stringent requirements on the timeliness of information delivery because outdated information can result in wrong decisions and lead to severe consequences \cite{2017Radio}.
In supporting delay-sensitive URLLC services as above, \textit{Age of Information} (AoI) is proposed as an effective and tractable metric to quantify the timeliness \cite{sunmodiano2019age}.
Unlike conventional metrics such as delay and throughput, AoI is assessed from the receiver's perspective, measuring the time elapsed since the latest received information update was generated.
Since AoI can capture the timeliness of information deliveries where traditional metrics cannot, the AoI-oriented network designs often generate unconventional (and sometimes counter-intuitive) insights as well as solutions. 
Consequently, AoI has attracted considerable attention to the research of the information update and transmission timeliness in the next generation URLLC system.

Early studies of AoI primarily focused on point-to-point scenarios \cite{2012Real,M2016On,2020The} and found that the corresponding AoI-oriented optimization design is different from that for conventional metrics.
Specifically, \cite{2012Real} provided a general method for calculating the average AoI, and found that the conventionally update policy, namely the zero-wait policy that maximizes throughput or minimizes latency, does not always lead to an optimal AoI.
And it observed that in fact, contrary to intuition, it is more desirable in many cases to wait for a certain amount of time at the transmitter side from an AoI optimum perspective.
\cite{M2016On} developed and compared the AoI performance under three different packet management schemes.
It is discovered that the one with the replacement protocol is the best among them. 
Considering that real-world applications have different sensitivity toward information staleness, the performance under nonlinear evolution of AoI was studied in \cite{2020The}.

In practice, information systems often consist of a large number of entities communicating via a shared spectrum. 
Due to the broadcast nature of the wireless medium, simultaneous transmissions from different nodes can interfere with each other.
Interference often incurs transmission collisions and failure deliveries, which can significantly impede the communication quality.
In response, a line of works introduced protocol models (a.k.a. conflict graphs) to characterize the phenomenon of transmission collisions caused by interference.
These models assume that within a certain geographic region, as long as there are source nodes transmitting simultaneously it will lead to failure.
Based on this model, several strategies are proposed to select active links for channel access, aiming to control the network interference and reduce the information age.
\cite{2020Opti} obtained the optimal, as well as suboptimal (but low-complexity) scheduling strategies for minimizing both average AoI and peak age under different source models.
In \cite{KadSinUys:18}, authors proposed four different scheduling strategies, including Greedy, Randomized, Max-Weight and Whittle’s Index policies, to optimize the performance of AoI, and they derived performance guarantees for these strategies.
However, these centralized scheduling methods generally require unified coordination and decisions, which are not applicable to scenarios where there exist massive end-user devices or applications in which traffic is highly bursty \cite{2021AgeofInformation}.
Therefore, distributed random access protocols, such as slotted ALOHA (SA) and carrier-sense multiple access (CSMA), attract considerable attention.
\cite{2020CSMA} provided closed-form expressions for the average AoI and average peak AoI under two different packet management schemes with and without preemption.
As the low-cost devices have low overhead budget and may not have carrier sensing capability, in which case SA-type protocol is preferable.
In \cite{2017Statusupdates}, the optimization of AoI using SA channel access strategy was studied, showing that its performance is inferior to that of round robin policy.
In view of this, some studies have modified SA, the work in \cite{2020Improving} proposed a deformation called \textit{Threshold ALOHA}, by which only when a source node's AoI exceeds a certain threshold will it be active and generate new information with a given probability. 
This channel access protocol can reduce the competition among sources to a certain extent, so as to improve the stale sources' probability of successful transmission and reduce the waste of transmission power.
In \cite{2020Closed}, an index was introduced to reflect the urgency of update, and the nodes were selected to access the wireless channel according to their index.

However, the conflict graph model oversimplifies the physical layer features by resorting to a binary judgment of the existence/non-existence of interference. In addition, it does not capture the critical attributes of a wireless system such as fading, path loss, power control, and co-channel interference.
As such, adopting the signal-to-interference-plus-noise ratio (SINR) or signal-to-interference ratio (SIR) model to characterize interference in the network becomes a more appealing method \cite{2017Birth}.
Based on this model, another line of work employ stochastic geometry as a tool to model the node position distribution as a point process, in order to account for those intricate effects from interference incurred by simultaneous transmissions.
In \cite{HuZhoZha:18}, the authors derived the upper and lower bounds of the cumulative distribution function (CDF) for the average AoI in the network by taking nodes' positions and their mutual interference into consideration.
\cite{2021SpatialDis} improved the bound on the spatial distribution of peak AoI and offered an exact bound on the successful transmission probability. 
The work in \cite{MankarTWC21} presented a stochastic geometric analysis of throughput and AoI in a cellular IoT network considering the spatial disparity in the AoI performance. The authors in \cite{2021spatiotemporal} provided expressions for AoI by using meta distribution, and their simulation results showed that the AoI performance is highly dependent on the spatial location, traffic load, and decoding threshold.
\cite{2021Understand} established a theoretical framework to analyze the impact of spatially interacting queues on AoI, which facilitates understanding network parameters' effect on the age performance.
And \cite{2022Spatio} extended the AoI analysis to networks where source nodes have unit-size buffers and operate under the last-come first-serve with replacement protocol, which is generally AoI-optimal. 
On the basis of such spatiotemporal models, a series of studies have been carried out to design schemes for performance optimization.
In \cite{YanArafaQue:19}, a decentralized channel access strategy was developed. Under this strategy, each node makes the AoI-optimal transmission decision according to the observation of its ambient communication environment.
An equation related to the surrounding parameters of each point was given in \cite{2022Locally}. 
By solving this equation, one can obtain each source's optimal status updating rate, adapting to its local transmission condition.
Additionally, utilizing the source nodes' local observation, \cite{2021powctl} designed a decentralized power control strategy by which nodes can adjust their transmit power individually to optimize the AoI performance.

Although these existing works have analyzed, as well as optimized, the AoI from various aspects, their designs are mainly pertaining to time slot-based system dynamics. 
In contrast, the effect of frame slotted ALOHA (FSA), an emerging technology that regularizes transmissions in the temporal domain and is now prevailing to IoT applications \cite{2021Stabi}, such as coordinating massive access in machine to machine (M2M) data collection networks \cite{2018M2M} and communications between readers and tags in radio frequency identification (RFID) systems \cite{2020RFID}, on information freshness remains largely unexplored.
Capitalized on the philosophy of FSA, this paper presents a new status updating protocol that improves age performance in a wireless network.               
Specifically, we organize a fixed number of consecutive time slots into a frame, and the source nodes determine to activate in each frame (or not) independently with a certain probability. 
If a source node decides to activate in a frame, it selects one time slot in the frame uniformly at random; upon the selected time slot, the source node generates a new update of status and immediately sends this information to its destination. 
We develop a theoretical framework that characterizes the performance of the proposed protocol. 
In particular, we model the spatial positions of the source and destination nodes as an ergodic and stationary point process.  
The sources update status information to their destinations using the FSA-based protocol. 
We consider an interference-limited scenario in which transmissions over a wireless link succeeds only if the SIR received at the destination exceeds a decoding threshold. 
We derive expressions for the average and quadratic AoI of a typical node by conditioning on the network topology.
Since the interference is affected by the spatial distribution of transmission links, we employ stochastic geometry to average the potential geographical patterns of the nodes, and derive closed-form expressions for the average and variance of AoI under two commonly used network models, i.e., the Poisson bipolar network and Poisson cellular network.
Leaning on the analytical results, we find that FSA provides a way to distribute sources into different time slots for update transmissions, thereby it can reduce the competition among sources in the network.
We show that FSA can always achieve a smaller average AoI than SA.
In addition, in the time slot of each frame after the update, since the source has no chance to transmit again, it will sleep to reduce power consumption.
During the period of preparing this paper, we found a very recent work that also studied the average AoI under FSA-based protocol \cite{2022FSA}.
Nonetheless, there are marked distinctions between our work and \cite{2022FSA}. 
Specifically, \cite{2022FSA} considered a single cell case in which multiple users transmit to a common access point.
And it adopted a protocol model to characterize the competition among users.
The authors of \cite{2022FSA} only analyzed the average AoI and gave a few elementary insights into the FSA status updating protocol.
In contrast, we consider a multi-cell setting, which is more practical, and use two specific network models that have been verified to well suit real-world scenarios \cite{2012Pair,2016Int}.
Moreover, we apply the SINR model, capturing more intricate features from a wireless system. 
We derive not only the average, but also variance of AoI, and the analysis can be extended to higher order moments or the \textit{Cost of Update Delay (CoUD)} \cite{2022nonlinear}. 
In addition, we have conducted a more comprehensive analysis by exploring network's age performance in various special-case scenarios, facilitating a thorough understanding to the impact of FSA updating protocol on AoI.
Our main contributions are summarized below.
\begin{itemize}
    \item We propose an FSA-based protocol for a set of source nodes to update their status information toward the intended destinations in a random access network. We develop a mathematical framework to evaluate the age performance of transmitters under such a protocol. Our model encompasses several key features of a wireless system, including channel fading, path loss, and interference. By fixing the network topology, we derive the (conditional) first and second moments of time-averaged AoI of a typical node, and verify the accuracy via simulations.
    \item When the nodes form a Poisson bipolar network, we obtain closed-form expressions for the average and variance of AoI of a typical node under the FSA-based status updating protocol. We compare the nodes’ age performance under FSA to those under the SA protocol and identify conditions under which FSA is instrumental in reducing the AoI. We also find that under the same updating frequency, converting an SA protocol into an FSA one always benefits the average AoI. Besides, the FSA-based protocol also avails the transmitters in reducing the transmission power consumption. 
    \item In the setting of a Poisson cellular network, we derive analytical expressions for the average and variance of AoI over a typical transmission link by accounting for effects of transmission protocol and power control strategy. The analysis allows us to quantify the control factors from signal power domain and interference domain, and their joint influence on the age performance. We also carry out several special case studies to garner useful insights. 
    \item Numerical results reveal that: \textit{i}) the FSA-based status updating protocol is beneficial when the spatial contention among transmitters is intense, i.e., the wireless links are frequently activated and densely deployed in space. In this situation, an optimal framesize exists that minimizes the average (or variance of) AoI;
    \textit{ii}) the gain of FSA is particularly pronounced when the network is densified. Specifically, when the deployment density increase by five folds, an FSA-based protocol can reduce the average and variance of AoI by two orders of magnitude compared to those under conventional SA-based protocol; and \textit{iii}) implementing FSA in conjunction with power control strategy can further benefit the network’s age performance, although the values of framesize and power control factor must be adequately tuned to achieve the optimal gain. 
\end{itemize}

The remainder of this paper is organized as follows.
Section~\ref{sec:sysmod} lays down the general system model and status updating protocol for this work.
By conditioning on the spatial layout, Section~\ref{sec:pre} shows the preliminary analysis framework.
Simulations are also provided to validate the accuracy of our results.
In Section~\ref{sec:bipolar}, we analyze the impact of FSA-based status updating protocol on the AoI performance in Poisson bipolar networks.
We derive closed-form expressions for the average and variance of AoI, and provide a series of discussions for insights. 
Similarly, Section~\ref{sec:cellular} explores the nodes' age performance in a Poisson cellular network. 
We also discuss the interplay between status updating protocol and power control policy in this section.
Finally, we conclude the paper in Section~\ref{sec:conclusion}.

\section{System Model}\label{sec:sysmod}
\begin{table} \label{table:notation}
\caption{Notation Summary
} \label{table:notation}
\begin{center}
\renewcommand{\arraystretch}{1.3}
\begin{tabular}{c  p{5.5cm} }
\hline
{\bf Notation} & {\hspace{2.5cm}}{\bf Definition}
\\
\hline
$\Phi_{ \mathrm{s} }$; $\lambda_{ \mathrm{s} }$ & Stationary and ergodic point process modeling the locations of sources; source spatial deployment density \\
$\Phi_{ \mathrm{d} }$; $\lambda_{ \mathrm{d} }$ & Stationary and ergodic point process modeling the locations of destinations; destination spatial deployment density \\
$\Phi$; $\lambda$ & Superposition of stationary and ergodic point processes $\Phi_{ \mathrm{s} }$ and $\Phi_{ \mathrm{d} }$, i.e., $\Phi \triangleq {\Phi_{ \mathrm{s} } } \cup {\Phi_{ \mathrm{d} } }$; superposition of spatial deployment density $\lambda_{ \mathrm{s} }$ and $\lambda_{ \mathrm{d} }$ in Poisson bipolar network, i.e., $\lambda=\lambda_{ \mathrm{s} }=\lambda_{ \mathrm{d} }$ \\
$P_i$; $\epsilon$  &  Transmit power of source node $i$; power control factor \\
$\eta$; $F$ & Packet update rate; framesize, i.e., the number of time slots contained in each frame \\
$x_i$  & Position of source $i$  \\                                                                                                                                                                                                                                                                                                                                                                                                                    
$R_i$; $D_i$ & Distance between source $i$ and its associated receiver; distance between source $i$ and the typical receiver \\
$r$; $\alpha$ & Distance of the typical transmitter-receiver pair; path loss exponent \\
$\nu_i$; $h_i$  & State indicator of source $i$; channel fading from transmitter $i$ to the typical receiver \\
$\theta$  & SINR decoding threshold \\
$\mu_0^\Phi$ & Transmission success probability of link $0$, conditioned on the point process $\Phi$ \\
$\bar{\Delta}_0$; $\sigma^2_{ \Delta_0 }$ & Average AoI over the typical link; variance of AoI over the typical link \\
\hline
\end{tabular}
\end{center}\vspace{-0.63cm}
\end{table}
In this section, we detail the setup of our network and the status updating protocol.
We also define the average and variance of AoI over a typical link. 
The main notations used throughout this paper are listed in the Table~\ref{table:notation}.

\subsection{Network Structure}
\subsubsection{Spatial Configuration}
We consider a wireless network deployed on the Euclidean plane, consisting of source and destination nodes. 
The spatial positions of the sources and their destinations are modeled by stationary and ergodic point processes, denoted by $\Phi_{ \mathrm{s} }$ and $\Phi_{ \mathrm{d} }$, respectively. 
Each source monitors an external process and transmits status updates to its intended destination. 
The status update of each source is encapsulated into information packets and transmitted over a shared spectrum.
Source node $i$ transmits an information packet with power $P_i$. 
We assume that the communication channel between any pair of nodes is narrowband, and it is affected by two attenuation components, i.e., the small-scale fading and large-scale path-loss, where the channel fading varies independently across time and space \cite{2012Stogeo}. 

\subsubsection{Temporal Attribute}
We consider a discrete-time system where the time is slotted into equal durations.
We assume the network is synchronized. 
We consider the updates of every source node are generated at the beginning of a time slot, whereas the transmission of an information packet takes up one time slot to finish. 
Furthermore, we package a fixed number of consecutive time slots into a frame, and the size of the frame is denoted by $F$. 
Since the time scale of fading and packet transmission is much smaller than that of the spatial dynamics, we assume the network topology is static, i.e., an arbitrary but fixed point pattern is realized at the beginning and remains unchanged in the subsequent time slots.

\subsubsection{Transmission Protocol}
Sources adopt an FSA-based transmission strategy synchronically.
Specifically, at the beginning of each frame, every source independently decides whether it will update in this frame or not, with probability $\eta$; if a source decides to update in this frame, it randomly selects a time slot according to a uniform distribution. 
Then, upon the transmission time slot, the source node generates a new update and immediately transmits the information packet to the destination. 
If the SIR at the destination exceeds a decoding threshold, the packet is successfully received; otherwise, the transmission fails.
By virtue of low-cost implementation, we do not employ a MAC protocol for the
re-transmission and/or acknowledgment of a reception. 
As such, every source will only send out an updated sample once. 



\begin{figure}[t!] 
  \centering{}

    {\includegraphics[width=0.98\columnwidth]{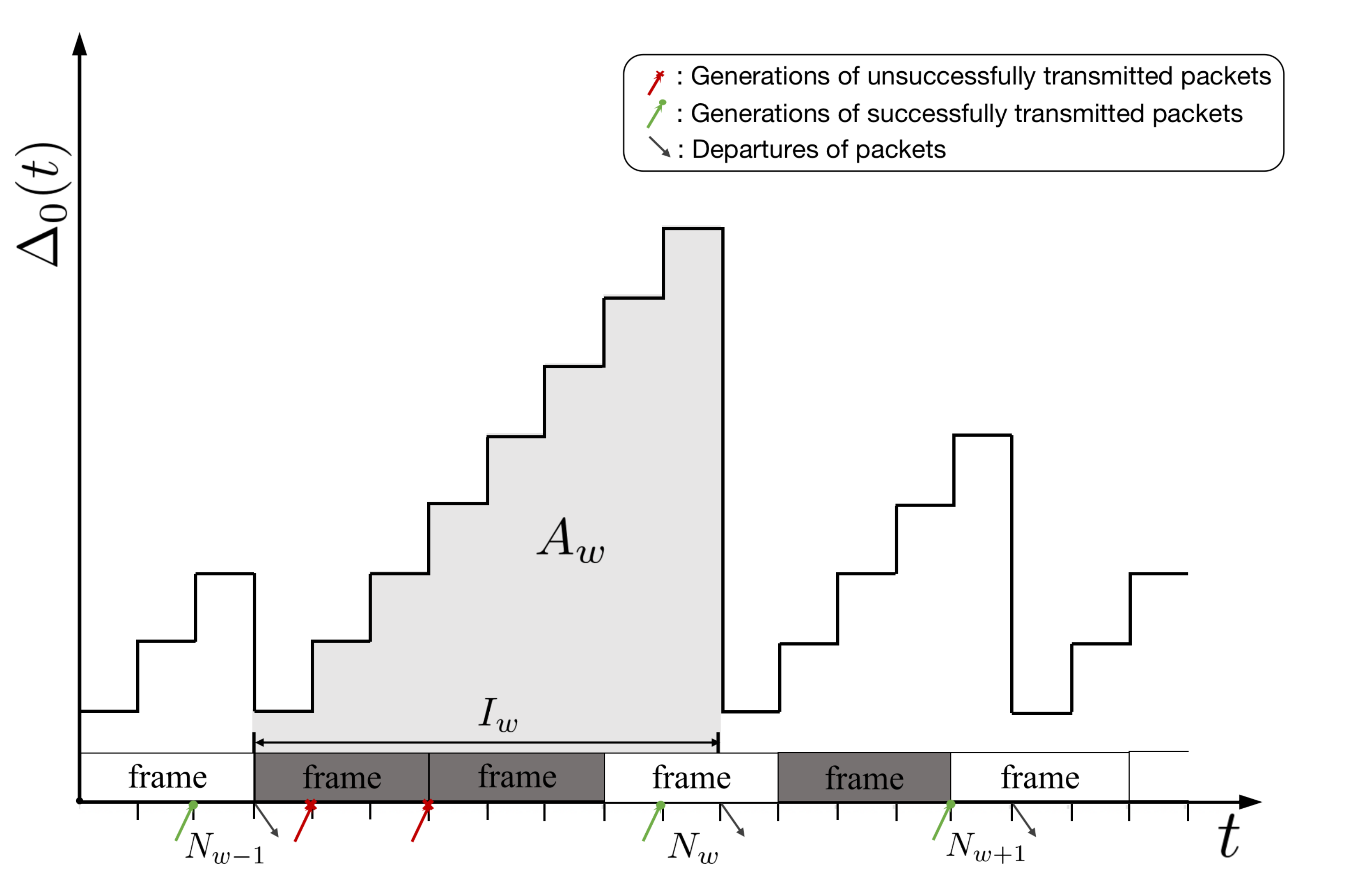}}

  \caption{An example of AoI evolution over the typical link under the FSA status updating protocol. The framesize is set as $F=3$.}
  \label{fig:FSA_AoI}
\end{figure}
\subsection{Performance Metrics}
Without loss of generality, we randomly select one transmission link in the network to be our \textit{typical link}, and denote the receiver's location as the origin.
Note that the performance of this link is statistically identical to the other links in the network, hence, it can serve as a representative one.

We consider AoI which captures the ``freshness" of information received at destinations.
AoI measures the time elapsed since the latest received data at a destination that is generated at the corresponding source. A formal definition of this metric is expressed as follows:
\begin{align} \label{equ:AoI}
    \Delta_0 (t) = t - G_0(t),
\end{align}
where $G_0(t)$ indicates the time-stamp of the most recently update received by the typical destination at time $t$. 
An example of AoI evolution under the proposed transmission protocol is illustrated in Fig.~\ref{fig:FSA_AoI}.

We employ the most representative metric, i.e., the average AoI over the typical link, to quantitatively evaluate the timeliness of information delivered in this network, which is defined as follows: 
\begin{align}
    \bar{\Delta}_0 = \lim_{T \to \infty }{\frac{1}{T}\sum_{t=1}^T \Delta_0(t)}.
\end{align}
In addition, we assess the variance of AoI {\footnote{It is noteworthy that the framework developed in this paper can be extended to study the network's age performance under more general cost functions by using similar approaches in \cite{2022nonlinear}. }} to investigate the reliability of network age performance. 
This metric is formally defined as follows:
\begin{align}
    \sigma^2_{ \Delta_0 }  = \lim_{T \to \infty }{ \frac{1}{T}\sum_{t=1}^T \left( \Delta_0(t) \right)^2 } - \left( \lim_{T \to \infty }{\frac{1}{T}\sum_{t=1}^T \Delta_0(t)} \right)^2.
\end{align}

\section{Preliminary Results} \label{sec:pre}
In this section, we derive analytical expressions for the first and second moments of AoI by fixing the network topology.
We verify these results by simulations. 

\subsection{Conditional AoI Statistics}

\subsubsection{SIR and Conditional Transmission Success Probability}
We consider an interference-limited scenario and adopt SIR to evaluate the transmission quality of wireless links. If the typical transmitter, situating at $x_0$, transmits a packet to its intended receiver at a time slot $t$, the received SIR can be written as:
\begin{align} \label{equ:SIR}
    \textrm{SIR}_{0,t} = \frac{ P_0 h_{0,t} L(r)^{-1}}{\sum_{i \neq 0}P_i h_{i,t} \nu_{i,t} L(\Vert x_i \Vert) ^{-1}},
\end{align}
where $x_i$ represents the position of transmitter $i$, $r=\Vert x_0 \Vert$ is the distance between the typical source-destination pair, $h_{i,t}$ denotes the channel fading from transmitter $i$ to the typical receiver, $\nu_{i,t} \in \{0,1 \}$ represents the state of transmitter $i$, where $\nu_{i,t}=1$ if transmitter $i$ is active and $\nu_{i,t}=0$ otherwise,
and $L: \mathbb{R}^+ \rightarrow \mathbb{R}^+$ is a monotonically non-decreasing function that characterizes the large-scale path loss.

Owing to the uncertainty in node positions, channel fading, and the activity of source nodes, the SIR is a random variable. 
A commonly used notion to characterize the statistical behavior of such a quantity is the \textit{conditional transmission success probability} \cite{haenggi2016meta}.  
Formally, given a point process $\Phi \triangleq {\Phi_{ \mathrm{s} } } \cup {\Phi_{ \mathrm{d} } }$, the conditional transmission success probability over the typical link is defined as:{\footnote{Since the point process is stationary and the sources activate independently from each other under the FSA-based protocol, the interference nodes also form a stationary point process. As such, we drop the time index $t$ from the subscript in the sequel. }} 
\begin{align} \label{equ:CndTXProb}
\mu_0^\Phi = \mathbb{P}(\mathrm{SIR_0}>\theta \mid \Phi),
\end{align}
where $\theta$ is the decoding threshold.

\subsubsection{Conditional AoI}
By fixing the spatial topology $\Phi$, dynamics of packet delivery over each wireless link can be regarded as a Bernoulli process that varies on a frame basis, where the active probability and transmission success probability are $\eta$ and $\mu_0^\Phi$, respectively.
Based on this abstraction, we derive the conditional average AoI in the following. 

\begin{theorem}
\textit{Conditioned on the spatial topology $\Phi$, the average AoI over the typical link is given as:
\begin{align} \label{equ:AoI}
    \mathbb{E}\left[\Delta_0 \vert \Phi \right]=\frac{F^2-1}{12F}\times \eta \mu_0^\Phi + \frac{F}{\eta \mu_0^\Phi}+\frac{1-F}{2}.
\end{align}
}
\end{theorem}
\begin{IEEEproof}
Please see Appendix~\ref{pro:The1}.
\end{IEEEproof}

Similarly, we can calculate the conditional quadratic AoI and it is provided by the next theorem.

\begin{theorem}
\textit{Conditioned on the spatial topology $\Phi$, the average quadratic AoI over the typical link is given as:
\begin{align} \label{equ:varience_AoI}
    \mathbb{E}\left[\Delta_0^2 \vert \Phi \right]=\frac{2F^2}{(\eta \mu_0^\Phi)^2} - \frac{F(2F\!\!-\!\!1)}{\eta\mu_0^\Phi}+\frac{F^2\!\!-\!\!1}{12F}  \eta\mu_0^\Phi+\frac{F(F\!\!-\!\!1)}{2}.
\end{align}
}
\end{theorem}
\begin{IEEEproof}
Please see Appendix~\ref{pro:The2}.
\end{IEEEproof}

Notably, $F$ not only affects AoI in an explicit manner, as we can see directly from expressions \eqref{equ:AoI} and \eqref{equ:varience_AoI}, but also implicitly affects the change of AoI through influencing $\mu_0^\Phi$. 
As such, the average and variance of AoI can be obtained by deconditioning \eqref{equ:AoI} and \eqref{equ:varience_AoI} with respect to point process $\Phi$, respectively, where different distributions of $\Phi$ lead to different results.
In the sequel, we will present two types of wireless network models, i.e., the Poisson bipolar network and Poisson cellular network, to examine the effect of FSA on the AoI performance. 
Before proceeding with a more detailed analysis, we will validate these theoretical results via simulations.

\begin{figure}[t!]
  \centering
  \subfigure[\label{fig:2a}]{\includegraphics[width=0.98\columnwidth]{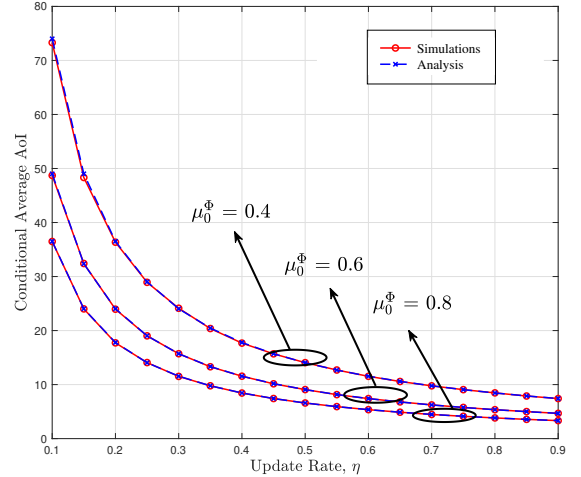}} ~
  \subfigure[\label{fig:2b}]{\includegraphics[width=0.98\columnwidth]{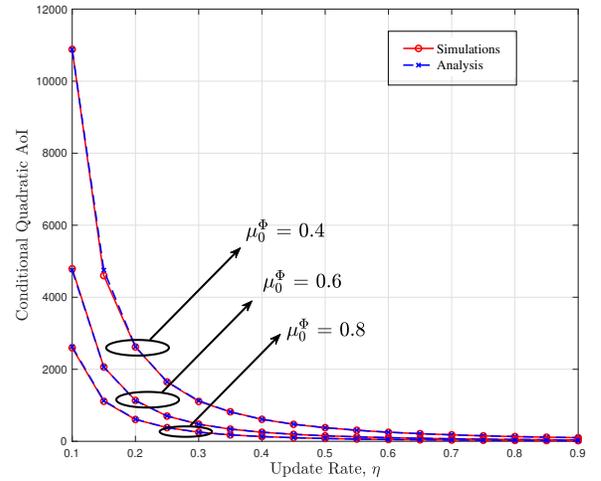}}
  \caption{Simulations versus analysis: ($a$) conditional average AoI and ($b$) conditional quadratic AoI, in which we vary the conditional transmission probability as $\mu_0^\Phi=0.4,0.6$, and $0.8$. 
  }  
  \label{fig:conditional_AoI}
\end{figure}
\subsection{Analysis Validation}

The simulations conducted in this part are dedicated to verifying the above analysis.
Specifically, we consider a source-destination pair and fix the conditional transmission probability. We set $F=3$ and run the simulation over $4 \times 10^6$ time slots.
We collect the AoI statistics and average them to obtain the final results.

In Fig.~\ref{fig:conditional_AoI}, we plot the average AoI and quadratic AoI as a function of update rate\footnote{In this work, the update rate is equivalent to the update probability or the sampling probability.}, by fixing the values of conditional transmission success probabilities. 
This figure shows that the simulations and analytical results are almost indistinguishable, which verifies our theoretical derivations.
We also observe that an increase in the (conditional) transmission success probability enhances the performance of the average AoI and average quadratic AoI. 

\section{Poisson Bipolar Networks} \label{sec:bipolar}
In this section, we analyze the effect of FSA on the AoI performance in a Poisson bipolar network. 
Such a model is motivated by the emerging interest in applications like Device-to-Device (D2D) networking, mobile crowd sourcing, and the Internet-of-Things (IoT), which do not require a centralized infrastructure, e.g., base stations or access points.

\subsection{Setting} \label{sec:bipo_set}
\begin{figure}[t!] 
  \centering{}

    {\includegraphics[width=0.97\columnwidth]{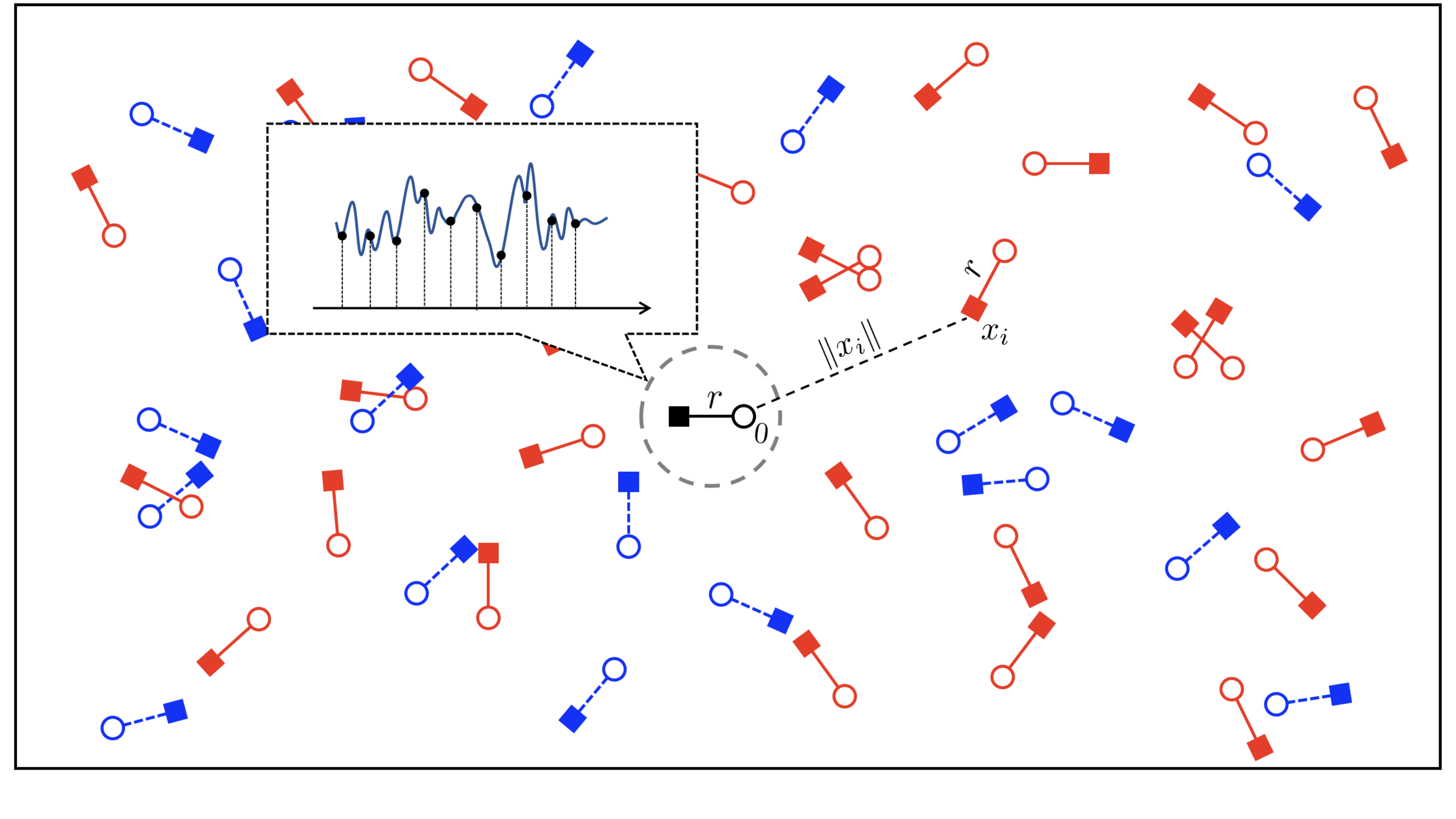}}

  \caption{A snapshot of the employed Poisson bipolar network, where the squares represent the sources, continuously sampling a sequence of status information and sending it to the destinations denoted by the circles. The solid black line is the typical link, the solid red lines represent other active links, and the dashed blue lines are inactive links.}
  \label{fig:Bipolar}
\end{figure}
In a \textit{Poisson bipolar network}, transmitters form a homogeneous Poisson point process (PPP) $\Phi_{ \mathrm{s} }$ of intensity $\lambda$.
Each transmitter has a dedicated receiver in a random orientation of a constant distance $r$. 
According to the displacement theorem \cite{BacBla:09}, the spatial layout of receivers $\Phi_{ \mathrm{d} }$ is also a homogeneous PPP of intensity $\lambda$.

We assume that every source node transmits at a constant power $P_{ \mathrm{tx} }$.
We also assume that the signal propagation is subjected to small-scale Rayleigh fading and standard path loss.
As such, SIR at the typical receiver can be written as:
\begin{align}\label{equ:SIR_PBN}
    \mathrm{SIR}_0^\mathrm{B}=\frac{ P_{ \mathrm{tx} } h_0r^{-\alpha}}{\sum_{i \neq 0} P_{ \mathrm{tx} } h_{i} \nu_{i} \Vert x_i \Vert ^{-\alpha}},
\end{align}
where $\alpha$ is the path-loss exponent.

Based on \eqref{equ:SIR_PBN}, we analyze the AoI performance under FSA and explore the interplay among the AoI metric and different network parameters in the following.

\subsection{Analysis}
Similarly to \cite{2015The}, we commence the AoI analysis by averaging out the effect of channel fading, which brings us to the following expression for the conditional transmission success probability. 
\begin{lemma} \label{lem:mu_0}
\textit{Given point process $\Phi$, the conditional transmission success probability over the typical link is:
\begin{align} 
    \mu_0^{\Phi} = \prod_{i \neq 0}\left(  1-\frac{\eta /F}{1+\Vert x_i \Vert ^{\alpha}/\theta r^\alpha} \right).
\end{align}
}
\end{lemma}
\begin{IEEEproof}
Please see Appendix~\ref{pro:Lem1}.
\end{IEEEproof}

Using this result, we can decondition $\mu^\Phi_0$ in \eqref{equ:AoI} and obtain the analytical expression for the average AoI.
\begin{theorem} \label{the:bipo_ave}
\textit{In a Poisson bipolar network, the average AoI of the typical link under FSA updating protocol is given by
\begin{align} \label{equ:bipo_average}
    \bar{\Delta}_0^ \mathrm{B} \!=\! \frac{F}{\eta}\exp\!{\Big(\frac{C\eta}{F} \big(1\!\!-\!\!\frac{\eta}{F}\big)^{\delta-1}}\Big)\!+\!\frac{(F^2\!\!-\!\!1)\eta}{12F}\exp\!{\Big(\!\!-\!\!\frac{C\eta}{F}\Big)}\!+\!\frac{1\!\!-\!\!F}{2},
\end{align}
where $\delta=2/\alpha$ and $C=\lambda \pi r^2 \theta^\delta \Gamma(1-\delta)\Gamma(1+\delta)$, with $\Gamma(\cdot)$ being the Gamma function \cite{2015The}.
}
\end{theorem}
\begin{IEEEproof}
Please see Appendix~\ref{pro:The3}.
\end{IEEEproof}

This theorem provides a closed-form expression that accounts for several key factors of a wireless system, i.e., the network topology, transmission protocol, and interference, on the average AoI. The following observations can be readily made from \eqref{equ:bipo_average}.

\subsubsection{When $F=1$}
The result in \eqref{equ:bipo_average} reduces to the classical average AoI under the SA protocol, given by
\begin{align} \label{equ:SA_average}
    \bar{\Delta}_0^\textit{SA}=\mathbb{E} \left[\frac{1}{\eta \mu_0^\Phi}\right]=\frac{1}{\eta}\exp{\left(C\eta (1-\eta)^{\delta-1}\right)}.
\end{align}
Note that the path loss exponent $\alpha$ generally satisfies $\alpha>2$ in practice, and consequently we have $\delta < 1$.
Then, following \eqref{equ:SA_average}, we can see that the average AoI is unbounded in regimes where the sources are generating updates in either extremely lazy (i.e., $\eta \rightarrow 0$) or excessively aggressive (i.e., $\eta \rightarrow 1$) manner. 
The former mainly ascribes to the absence of new updates at the sources; the latter due to potential interferers located in geographical proximity to the typical receiver, which impedes the transmissions and hinders any possible packet delivery.
In contrast, \eqref{equ:bipo_average} indicates that incorporating FSA to the status updating protocol can effectively abbreviate such severe interference issue (note that for $F>1$, the average AoI is always bounded when $\eta \rightarrow 1$).

\subsubsection{When $F>1$}
In this case, let us set $\beta=\frac{\eta}{F}$ and term it as the \textit{effective updating rate} under FSA. 
Then, we can rewrite \eqref{equ:bipo_average} as follows:
\begin{align}\label{equ:beta_form_AveAoI}
    \bar{\Delta}_0^\mathrm{B}= \bar{\Delta}_0^\textit{SA}(\beta)+\underbrace{\frac{F^2-1}{12}\beta \exp{\Big(-C\beta \Big)} + \frac{1 - F}{2} }_{Q_1},
\end{align}
where $\bar{\Delta}_0^\textit{SA}(\beta)=\frac{1}{\beta}\exp{(C\beta (1-\beta)^{\delta-1}})$ denotes the average AoI of the typical link under SA with updating rate $\beta$. 
From \eqref{equ:beta_form_AveAoI}, we note that if $Q_1$ is non-negative,
introducing frame structures into SA is not indispensable since one can always obtain a smaller average AoI without FSA by adjusting the update rate $\beta$ under SA.  
Therefore, FSA-based protocol is instrumental in reducing AoI only when $Q_1<0$. 

An interesting observation is that regardless of the interference level, imposing a frame structure on the SA protocol always improves the network age performance. 
Below we present two approaches to bridge the SA and FSA protocols. 
To facilitate exposition, we use $\eta_{\mathrm{SA}}$ to denote the state update rate under SA protocol and $\eta_{\mathrm{FSA}}$ to denote that under FSA.
We can convert any SA at $\eta_{\mathrm{SA}}$ to:

a) FSA with $\eta_{\mathrm{FSA}}=2\eta_{\mathrm{SA}}$, $F=2$.
Note that in this scenario, the effective updating rate of the FSA protocol is $\beta = \eta_{\mathrm{SA}}$, which is identical to that of SA. 
Consequently, we can compute $Q_1$ as follows:
\begin{align}
    Q_1 = \frac{\eta_{\mathrm{FSA}}}{8} \exp{\Big(-C\frac{\eta_{\mathrm{FSA}}}{2} \Big)} - \frac{1}{2}
    < \frac{\eta_{\mathrm{FSA}}}{8} - \frac{1}{2}
    <0.
\end{align}

b) FSA with $\eta_{\mathrm{FSA}}=1$, $F=\frac{1}{\eta_{\mathrm{SA}}}$, if $\frac{1}{\eta_{\mathrm{SA}}}$ is an integer.
In this scenario, $\beta = \frac{1}{F}$, which is also identical to that of SA. 
To demonstrate that FSA outperforms SA, we rewrite $Q_1$ by the following:
\begin{align}
    Q_1 &= \frac{F^2\!-\!1}{12 F}  \exp{\Big(\!-\!\frac{C}{F} \Big)} + \frac{1 \!-\! F}{2} 
    \nonumber\\
    &< \frac{F^2\!-\!1}{12 F}  + \frac{1 \!-\! F}{2} 
    = - \frac{ (5F-1) (F - 1)}{12 F}  <0.
\end{align}
Both of the above designs can transform the commonly used SA updating protocol into an FSA protocol and achieve better AoI performance.
This is because FSA not only decreases mutual interference amongst the transmitters, but also equalizes the updating intervals of each source node, thereby reducing the AoI \cite{sunmodiano2019age}.

\subsubsection{The optimal $F$}
By applying the inequality of arithmetic and geometric means to \eqref{equ:bipo_average}, we can derive a lower bound to the average AoI over the typical link as:
\begin{align} \label{equ:low_boun}
    \bar{\Delta}_0 &\geq 2 \sqrt{\frac{(F^2-1)\eta \mu_0^\Phi}{12F} \times \frac{F}{\eta \mu_0^\Phi}} +\frac{1-F}{2} 
    \nonumber\\
    &= 2\sqrt{\frac{F^2-1}{12}}+\frac{1-F}{2}.
\end{align}

This inequality indicates that merely increasing $F$ does not always benefit the AoI performance.
In consequence, we shall adequately choose the framesize in accordance with the network parameters to achieve the optimal operation regime. 

\begin{figure}[t!]
  \centering{}

    {\includegraphics[width=0.98\columnwidth]{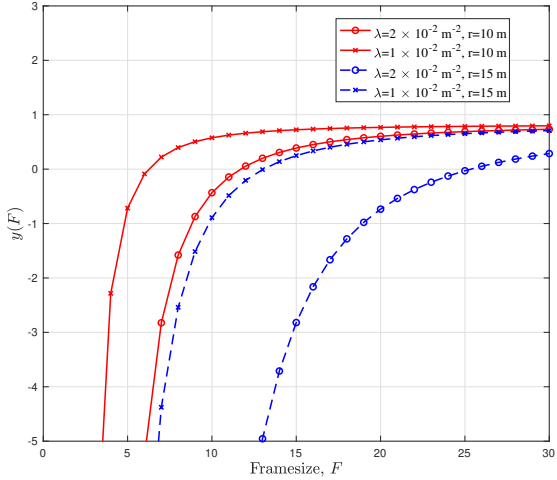}}

  \caption{ A plot of function $y(F)$ verse framesize under different spatial deployment parameters, in which we set $\alpha=3.5$, $\eta=0.8$, $\theta=0 \, \mathrm{dB}$, and vary $\lambda$ as $\lambda=2 \times 10^{-2}$, $\lambda=1 \times 10^{-2} \, \mathrm{m}^{-2}$, $r$ as $r=10$, $r=15 \, \mathrm{m}$. }
  \label{fig:optimal_F}
\end{figure}

With this understanding, we fix other parameters and explore the optimal $F$ that minimizes the average AoI.
Specifically, we relax the constraint of $F$ being an integer, take the derivative of $\bar{\Delta}_0^\mathrm{B}$ with respect to $F$, and assign $\frac{\mathrm{d}\bar{\Delta}_0^\mathrm{B}}{\mathrm{d}F}=0$.
Then, we can obtain the theoretical optimal $F$ (denoted by $F^*$) by solving $y(F)=0$, where $y(F)$ is expressed as:
\begin{align} \label{equ:bipo_opF}
    y(F)&=\left[\frac{1}{\eta}+\frac{C(1\!\!-\!\!\frac{\eta}{F})^{\delta-2}}{F^2} (\eta \delta-F) \right] \exp\!{\Big(\frac{C\eta}{F} \big(1\!\!-\!\!\frac{\eta}{F}\big)^{\delta\!-\!1}}\Big)
    \nonumber\\
    &+\frac{(F^2\!\!+\!\!1)F+(F^2\!\!-\!\!1)C\eta}{12F^3} \eta \exp\!{\Big(\!-\!\frac{C\eta}{F}\Big)}-\frac{1}{2}.
\end{align}
Note that the root of this function can be attained efficiently by popular software such as Matlab.
Based on the parameter setting as per Section \uppercase\expandafter{\romannumeral4}-C, we give a plot in Fig.~\ref{fig:optimal_F} on what $y(F)$ looks like.
This figure shows that $y(F)$ increases monotonically, and if the solution to $y(F)=0$ exists, it must be unique. 
Moreover, we shall assign the optimal framesize $F$ as the one between $\lceil F^* \rceil$ and $\lfloor F^* \rfloor$ that makes the $\bar{\Delta}_0$ smaller, where $\lceil \cdot \rceil$ and $\lfloor \cdot \rfloor$ denote the ceil and floor functions, respectively.
On the other hand, if $y(F)=0$ does not have a solution, we set $F^*=1$.

\subsubsection{Spatial throughput}
Another commonly used metric in the optimization of network deployment is spatial throughput, given by \cite{2011Tractable}: 
\begin{align} \label{equ:throughput}
    \Theta &\triangleq \frac{\eta}{F} \mathbb{P}( \mathrm{SIR}_0 > \theta ) \log(1 + \theta)
    \nonumber\\
    &= \frac{\eta}{F}\mathbb{E}[\mu_0^\Phi] \log(1 + \theta)  
    \nonumber\\
    &= \frac{\eta}{F} \exp{\left(-\frac{C\eta}{F}\right)} \log(1 + \theta),
\end{align}
where $\mathbb{E}[\mu_0^{\Phi}] = \exp(-C \frac{\eta}{F})$ follows from \eqref{equ:bipo_1}. 
This quantity measures the average rate of successful information delivery over the typical link. 
It is natural to infer that maximizing spatial throughput also optimizes AoI, because a high throughput enables packets to be delivered swiftly, giving “fresher” information at the receiver.
However, comparing \eqref{equ:bipo_average} and \eqref{equ:throughput}, we can see that jointly tuning the update rate $\eta$ and frame size $F$ to obtain a higher throughput, i.e., making the term $\frac{\eta}{F}\exp{\left(-\frac{C\eta}{F}\right)}$ as large as possible, does not necessarily lead to a smaller AoI.

\subsubsection{Transmission power consumption}
According to some recent studies on energy efficiency, sleep-scheduling strategies can effectively reduce power consumption and optimize AoI performance \cite{2022Energy,2021Low}. 
Considering that FSA also has an implicit ``sleep scheduling" characteristic, we investigated the effectiveness of FSA in promoting energy efficiency in addition to optimizing AoI performance.
More specifically, conditioned on the spatial topology $\Phi$, in each frame, let us denote by $U$ (resp. $\bar{U}$) the event that the typical transmitter does (resp.  does not) generate a new update, and $S$ (resp. $\bar{S}$) that the typical receiver does (resp.  does not) receive a successful update.
Moreover, we denote $E_0(w)$ as the energy consumed by the typical source node since the ($w-1$)-th successful transmission to the $w$-th one.
Then, we can compute the transmission power consumption over the typical link as:
\begin{align} \label{equ:TX_power_FSA}
     \mathbb{E}[P_0 \vert \Phi]
     &=\lim_{W \to \infty}\frac{\sum_{w=1}^W E_0(w)}{\sum_{w=1}^W I_w}
     \nonumber\\
     &=\frac{\mathbb{P}(U \vert \bar{S})(\mathbb{E}[X]-1)+1}{\mathbb{E}[I]} P_{ \mathrm{tx}}
     = \frac{\eta P_{ \mathrm{tx}}}{F},
\end{align}
where $I$ is the interval between two consecutive updates received, $\mathbb{P}(\bar{S})=1-\mathbb{E}[\mu_0^\Phi]$, and $\mathbb{P}(U)=\eta$.
Following \eqref{equ:TX_power_FSA}, we can clearly see that merging time slots into frames is capable of reducing the transmission power consumption.

\setcounter{equation}{\value{equation}}
\setcounter{equation}{18}
\begin{figure*}[t!]
\begin{align} \label{equ:vari_bipo}
    \sigma^2_{ \Delta^\mathrm{B}_0 }&=\frac{2F^2}{\eta^2} \exp{\left( -C\sum_{k=1}^\infty (k+1) {\delta-1 \choose k-1}\big(-\frac{\eta}{F} \big)^k \right)}
    - \frac{F^2}{\eta^2}\exp{\left(\frac{2C\eta}{F}\big(1-\frac{\eta}{F}\big)^{\delta-1}\right)} 
    - \frac{(F^2-1)^2\eta^2}{144F^2}\exp{\left(-\frac{2C\eta}{F}\right)}
    \nonumber\\
    &-\frac{F^2-1}{6}\exp{\left( \frac{C\eta}{F}\Big( \big(1-\frac{\eta}{F} \big)^{\delta-1}-1\Big) \right)}-\frac{F^2}{\eta}\exp{\left(\frac{C\eta}{F}\big(1-\frac{\eta}{F} \big)^{\delta-1}\right)}
    +\frac{(F^2-1)\eta}{12}\exp{\left(-\frac{C\eta}{F} \right)}+\frac{F^2+2F-3}{4}
\end{align}
\setcounter{equation}{\value{equation}}{}
\setcounter{equation}{19}
\centering \rule[0pt]{18cm}{0.3pt}
\end{figure*}
\setcounter{equation}{19}

In addition to computing the average AoI, we can also derive the analytical expression for the variance of AoI by deconditioning $\mu_0^\Phi$ in \eqref{equ:varience_AoI} to obtain $\mathbb{E}[\Delta_0^2]$ and then following similar calculations as the above.
\begin{theorem} \label{the:bipo_var}
\textit{In a Poisson bipolar network, the variance of AoI over the typical link under FSA updating protocol can be expressed as \eqref{equ:vari_bipo} at the top of this page.
}
\end{theorem}
\begin{IEEEproof}
Please see Appendix~\ref{pro:The4}.
\end{IEEEproof}

When $F=1$, this result reduces to the variance of AoI under the SA protocol.
We provide it in the following as a complement to the current studies of AoI in random access networks.

\begin{corollary}
\textit{In a Poisson bipolar network, the variance of AoI over the typical link under SA updating protocol is given by
\begin{align}
    &\sigma^2_{ \Delta^\textit{SA}_0 }
    =\mathbb{E}\left[\frac{2}{(\eta \mu_0^\Phi)^2}-\frac{1}{\eta \mu_0^\Phi}\right]-\mathbb{E}\left( \left[ \frac{1}{\eta \mu_0^\Phi}\right] \right)^2
    \nonumber \\
    &=\frac{2}{\eta^2}\exp{\left( -C\sum_{k=1}^\infty (k+1) {\delta-1 \choose k-1}(-\eta)^k \right)}
    \nonumber\\
    &- \frac{1}{\eta^2}\exp{\left(2C\eta(1-\eta)^{\delta-1}\right)} 
   -\frac{1}{\eta}\exp{\left(C\eta(1-\eta)^{\delta-1}\right)}.
\end{align}}
\end{corollary}

When $F>1$, we follow a similar analysis method to the average AoI and rewrite \eqref{equ:vari_bipo} in terms of effective update rate, as follows:
\begin{align} \label{equ:beta_form_VarAoI}
    &\sigma^2_{ \Delta^\mathrm{B}_0 }=\sigma^2_{ \Delta_0^\textit{SA} }(\beta)
    \nonumber\\
    &\underbrace{-\! \bigg(\! \frac{F^2\!\!-\!\!1}{12 \beta} e^{\!-\!\frac{C}{\beta}} \!-\! \frac{F}{2} \bigg)^{\!2}
    \!\!-\!\! \frac{F^2 \!\!-\!\! 1}{6} e^{ \frac{C\eta}{F}\Big( \big(1 \!-\! \frac{\eta}{F} \big)^{\delta\!-\!1} \!-\! 1\Big) }
    \!\!+\!\! \frac{2F^2\!\!+\!\!2F\!\!-\!\!3}{4} }_{Q_2},
\end{align}
where $\sigma^2_{ \Delta_0^\textit{SA} }(\beta)$ denotes the variance of AoI over the typical link under SA protocol with update rate $\beta$.
According to this result, we note that the FSA protocol is preferable to SA only when $Q_2<0$.

\subsection{Numerical Results} \label{sec:bipo_num}
Based on the analytical results, this part shows the AoI performance under different network operation regimes.
Unless otherwise specified, we use the following parameters: $\alpha=3.5$, $\theta=0 \, \mathrm{dB}$.

\begin{figure}[t!]
  \centering
  \subfigure[\label{fig:3a}]{\includegraphics[width=0.98\columnwidth]{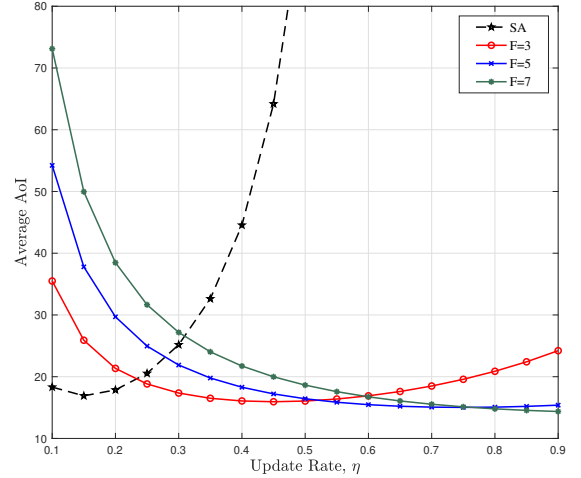}} ~
  \subfigure[\label{fig:3b}]{\includegraphics[width=0.98\columnwidth]{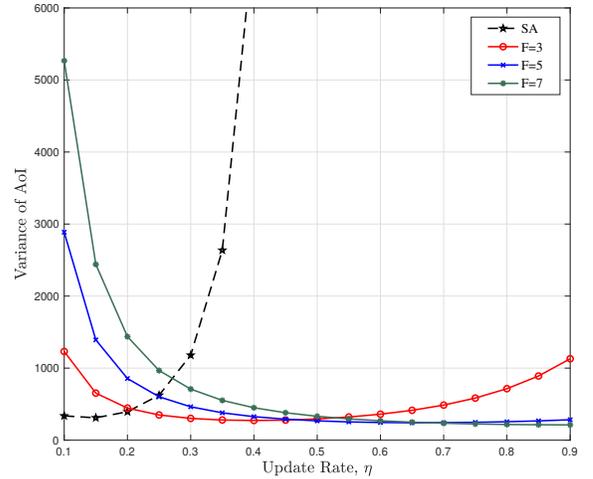}}
  \caption{AoI performance versus the update rate under different framesizes $F$: ($a$) average AoI and ($b$) variance of AoI, in which we set $\lambda=1 \times 10^{-2} \, \mathrm{m}^{-2}$, $r=10 \, \mathrm{m}$, and vary the framesize as $F=1$ (i.e. SA), $3$, $5$, and $7$.}  
  \label{fig:bipo_eta}
\end{figure}
Fig.~\ref{fig:bipo_eta} illustrates the average, as well as variance, of AoI as a function of the status updating rate, respectively, under different framesizes $F$.
From Fig.~\ref{fig:3a}, we observe that for the considered situations, there exists an optimal update rate that minimizes the average AoI. This mainly arises from the tradeoff between generating fresh information at the sources and maintaining interference at a relatively low level across the network. 
Moreover, we notice that when $\eta$ is relatively low, SA attains a smaller average AoI than FSA. 
Because, in this case, the interference is mild, and the sources shall not wait for the entire frame to generate a new update.
However, when $\eta$ increases, the average AoI under SA (i.e., $F=1$) grows rapidly. In comparison, those under FSA with $F \geq 2$ remain at relatively small values for a wide range of $\eta$. 
The reason attributes to the fact that increasing the update rate can activate more transmission links. 
As a result, interference among source nodes becomes destructive, leading to the consequence of many transmission failures. 
In contrast, FSA implicitly regularizes the nodes' transmission patterns by imposing a frame structure on their active periods. 
Notably, even if two sources are situated at close proximity in space, choosing a frame with size $F \geq 2$ can dramatically decrease the chance that these nodes select the same time slot for sending out the update information and result in the collision of their transmitted packets. 
Consequently, we can see that when $F$ is relatively large (i.e., $F=7$), the average AoI declines steadily as we increase the update rate. 
Because when $F$ is large, the nodes have more opportunities to pick different time slots for generating updates (and transmitting them).
As such, the transmission benefits from low interference; hence, the more frequent the generation of the updates, the fresher the received information. We observe similar phenomena from Fig.~\ref{fig:3b}, except that ($i$) the variance of AoI is more sensitive to the change of update rates; and ($ii$) therefore, the optimal $\eta$ that minimizes the variance of AoI is different from that of the minimum average AoI.
\textit{Nonetheless, it is noteworthy that using FSA for status updating in a large-scale wireless network not only reduces the average AoI but, more importantly, flats out the variations in the variance of AoI, which is crucial for stabilizing the network operation}.

\begin{figure}[t!]
  \centering
  \subfigure[\label{fig:4a}]{\includegraphics[width=0.98\columnwidth]{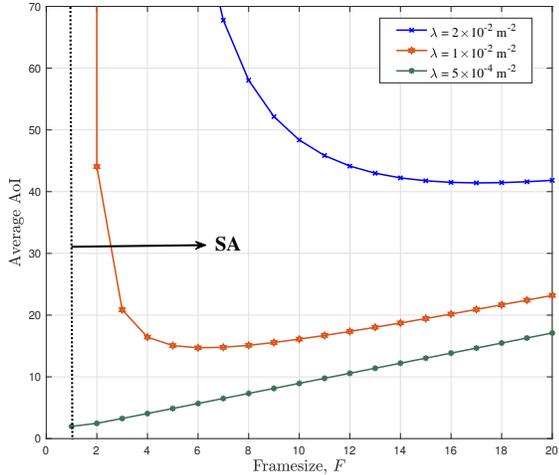}} ~
  \subfigure[\label{fig:4b}]{\includegraphics[width=0.98\columnwidth]{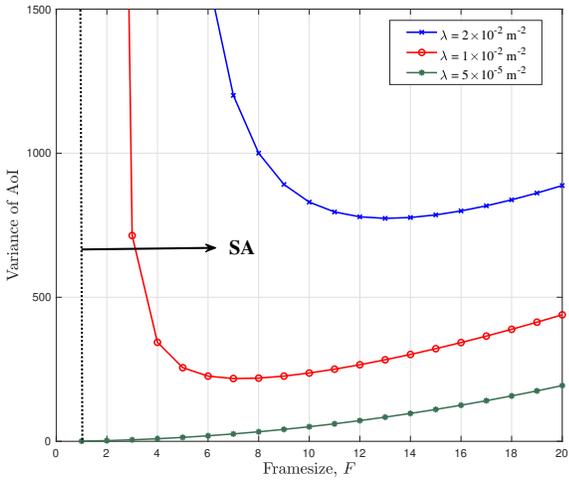}}
  \caption{ AoI performance versus framesize under different deployment densities $\lambda$: ($a$) average AoI and ($b$) variance of AoI, in which we set $\eta = 0.8$, $r=10 \, \mathrm{m}$, and vary the deployment densities as $\lambda=2 \times 10^{-2} $, $1 \times 10^{-2} $, $5 \times 10^{-3} $, and $5 \times 10^{-5}  \, \mathrm{m}^{-2}$.  }  
  \label{fig:bipo_F}
\end{figure}

\begin{figure}[t!]
  \centering
  \subfigure[\label{fig:4a}]{\includegraphics[width=0.98\columnwidth]{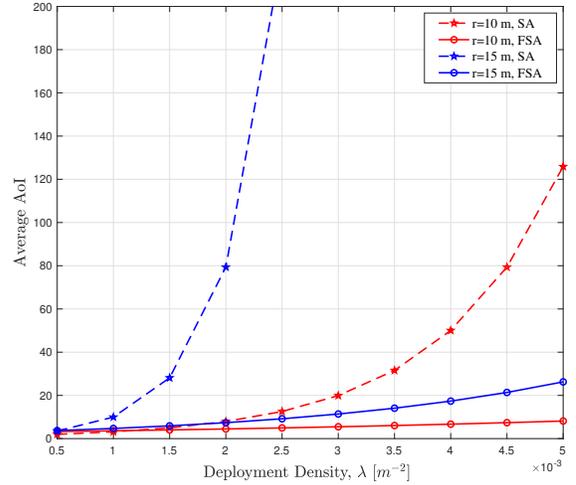}} ~
  \subfigure[\label{fig:4b}]{\includegraphics[width=0.98\columnwidth]{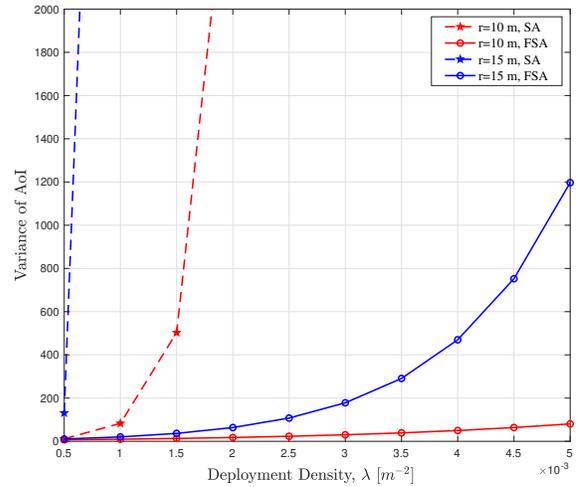}}
  \caption{ Comparison of AoI performance for deployment density variation under SA and FSA strategies. First, we fixed $\eta = 0.8$ in SA and $(\eta,F)=(0.8,3)$ in FSA, then plot: ($a$) average AoI and ($b$) variance of AoI.  }  
  \label{fig:bipo_lam}
\end{figure}

Recognizing that the frame size of FSA has a remarkable influence on the AoI performance, we plot the average and variance of AoI as functions of $F$ in Fig.~\ref{fig:bipo_F}.
The figure unveils that depending on the network configuration, there may (or may not) exist an optimal $F$ that minimizes the average/variance of AoI.
This is because, while enlarging $F$ can mitigate the conflicts among nodes' transmissions, which in turn increases the transmission success probability, it also prolongs the duration that each source generates a new update. 
To this end, we can see that the framesize strikes a delicate balance between the information freshness at the sources and the interference level across the network. In addition, we note that although increasing the deployment density will inevitably raise up the average and variance of AoI, adopting FSA alleviates such deterioration by effectually dwindling the transmission collisions. 
Furthermore, we notice that as $\lambda$ increases, the optimal $F$ goes up correspondingly, entailing more stringent regularization to curb competition among the transmitting nodes.
On the other hand, when the spatial deployment is sparse (e.g., $\lambda = 2 \times 10^{-2}$ in this case), interference is mild, and the FSA protocol loses its efficacy. Actually, it may perform worse than the SA protocol under such a circumstance.

Fig.~\ref{fig:bipo_lam} depicts the AoI statistics, i.e., the average and variance, as functions of the deployment density. 
In this example, we fix the framesize and update rate as $F=3$ and $\eta=0.8$, respectively. 
By comparing the AoI performance under SA and FSA, we find that networks under FSA protocol attain a remarkable reduction in the average and variance of AoI as the infrastructure is densified. 
More concretely, it can be seen from Fig.~\ref{fig:4a} that under a moderate source-destination distance $r=10$, when the spatial density increases by five folds, the average AoI under SA goes up by more than two orders of magnitude; in comparison, the increase in the average AoI under FSA is almost unnoticeable. 
Such a difference becomes more prominent when $r$ increases from 10 to 15, where the average AoI under SA surges (almost) without limit, while that under FSA only rises fractionally. 
On the other hand, Fig.~\ref{fig:4b} demonstrates that FSA is also effective in decreasing the variance of AoI--with a more pronounced gain than the average AoI.
These phenomena are in line with our previous analysis, that when the source nodes update fast and are densely distributed, FSA protocol performs better than SA (it corresponds to that $Q_1$ from (13) and $Q_2$ from (22) are both negative).
Under this circumstance, increasing $\lambda$ and $r$ enlarges $C$, leading to increases in the absolute values of $Q_1$ and $Q_2$, which widen the gap between SA and FSA.

\section{Poisson Cellular Networks} \label{sec:cellular}
This section explores the age performance in the setting of cellular networks. 
Under this model, multiple source nodes transmit status update information to a common destination. 
We derive analytical expressions for the considered AoI metrics, and investigate the interplay between FSA protocol and the intra-cell spectral competition, inter-cell interference, and the power control on the AoI performance. 

\subsection{Setting} \label{sec:cell_set}

\begin{figure}[t!] 
  \centering{}

    {\includegraphics[width=0.98\columnwidth]{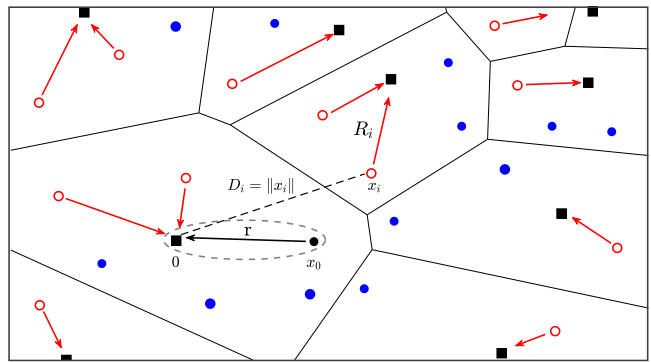}}

  \caption{An example of the Poisson cellular network, where the sources and destinations are denoted by the squares and circles, respectively. The solid black line is the typical link, the solid red lines represent other active links, and the solid blue circles are inactive sources with no connection.}
  \label{fig:Cellular}
\end{figure}

We consider the uplink of a Poisson cellular network, as depicted in Fig.~\ref{fig:Cellular}, where spatially distributed sensors need to update status information to their targeted data fusing centers. 
The data fusing centers are deployed according to a homogeneous PPP $\Phi_\mathrm{d}$ with spatial density $\lambda_\mathrm{d}$.
The sensors are scattered as an independent homogeneous PPP $\Phi_\mathrm{s}$ of intensity $\lambda_\mathrm{s}$.
Every sensor associates with the closest data fusing center in geographical space.

In the context of cellular networks, distances between different transmitter-receiver pairs vary significantly, leading to a crucial impact from path loss on the signal attenuation.
In view of this challenge, we consider each sensor adopts a power control strategy for its transmission.
Specifically, we denote by $\epsilon \in [0,1]$ and $R_i$ the power control factor and the distance between the $i$-th sensor located at $x_i$ to its associated data fusing center, respectively; then, the transmit power of sensor $i$ is $P_\mathrm{tx} R_i^{\alpha \epsilon}$.
Accordingly, the SIR of the typical receiver can be written as:
\begin{align} \label{equ:PCN_SIR}
    \mathrm{SIR}_0^\mathrm{C}=\frac{ P_\mathrm{tx} h_0 r^{\alpha(\epsilon-1)}}{\sum_{i \neq 0} P_\mathrm{tx} h_{i} \nu_{i} R_i^{\alpha \epsilon} \Vert x_i \Vert ^{-\alpha}}.
\end{align}

Using this expression, we will analyze the average and variance of AoI under FSA in the following section and investigate the AoI performance under different power control strategies.

\subsection{Analysis}
Similar to the previous section, we commence our analysis with deriving an initial expression for the conditional transmission success probability. 
Specifically, by substituting \eqref{equ:PCN_SIR} into \eqref{equ:CndTXProb} and averaging out the randomness from channel fading and interferers' active states, we have: 
\begin{align}
    \mu_0^{\Phi} &= \mathbb{P}\left( \mathrm{SIR}_0^\mathrm{C} > \theta | \Phi \right)
    = \prod_{i \neq 0}\left(  1-\frac{\eta /F}{1+ \frac{\Vert x_i \Vert ^{\alpha}}{\theta R_i^{\alpha \epsilon} r^{\alpha(1-\epsilon)}}} \right).
\end{align}

Based on this result, we can decondition $\mu_0^\Phi$ in \eqref{equ:AoI} and obtain the expression of the average AoI over the typical link. 
However, owing to effects of power control, the spatial locations of interference nodes is a non-stationary point process where the intensity depends on the distance between an interferor to the typical data fusing center,  hindering the derivation of an exact analysis. 
In light of this, we adopt the approximations developed in \cite{2016User} for the probability density function (pdf). 
According to \cite{2016User}, the pdf of the distance, $R_i$, between a generic sensor $i$ and its associated data fusing center can be approximated as 
\begin{align}
    f_r(u)&=\frac{5}{2} \lambda_\mathrm{d} \pi u \exp{\left(-\frac{5}{4} \lambda_\mathrm{d} \pi u^2 \right)}.
\end{align}
The distance between sensor $i$ and the typical data fusing center, which located at the origin, is given by 
\begin{align}
    f_x(u)&=2\pi\lambda_\mathrm{d} u \exp{(-\lambda_\mathrm{d} \pi u^2)}.
\end{align}
Due to the association policy, we know that $R_i$ and $D_i$ are not independent. In fact, $R_i$ cannot be larger than $D_i$, since otherwise sensor $i$ will associate to another data fusing center. 
We formalize this correlation by using the conditional distribution function given by:
\begin{align}
    f_r(u \vert D_i)=\frac{\frac{5}{2} \lambda_\mathrm{d} \pi u \exp{(-\frac{5}{4}\lambda_\mathrm{d} \pi u^2)}}{1-\exp{(-\frac{5}{4}\lambda_\mathrm{d} \pi D_i^2)}}.
\end{align}

Armed with the above preparation, we are now ready to present the analytical expression for the average AoI.
\begin{theorem}\label{thm:Cellular_AveAoI}
\textit{In a Poisson cellular network, the average AoI over the typical link under FSA updating protocol is given by
\begin{align} \label{equ:cell_ave}
    \bar{\Delta}_0^\mathrm{C} &=\!\! \int_0^\infty \!\! e^{-z} \Bigg( \frac{F^2\!\!-\!\!1}{12F}\eta \exp{\Big(\!\!- \frac{\lambda_\mathrm{s}}{\lambda_\mathrm{d}} \frac{\eta}{F} g_\theta(1,z;\alpha,\epsilon) \Big)}
    \nonumber\\
    &+\frac{F}{\eta}\exp{\Big(\frac{\lambda_\mathrm{s}}{\lambda_\mathrm{d}} \frac{\eta}{F} g_\theta(1-\frac{\eta}{F},z;\alpha,\epsilon) \Big)}\Bigg)\mathrm{d}z+\frac{1\!-\!F}{2},
\end{align}
where $g_\theta(m,z;\alpha,\epsilon)$ is given as: 
\begin{align}
    g_\theta(m,z;\alpha,\epsilon) \!= \frac{4}{5}z^2 \!\!\! \int_0^\infty \!\!\! \int_0^1 \! \frac{q \exp{(-zqs)} \mathrm{d}s \mathrm{d}q}{(m \!+\! \frac{q^{\alpha(1-\epsilon)/2}}{\theta s^{\alpha \epsilon/2}})(1-e^{-zq})}.
\end{align}
}
\end{theorem}
\begin{IEEEproof}
Please see Appendix~\ref{pro:The5}.
\end{IEEEproof}

The theorem above captures the interplay amongst the update rate, power control, and interferes' spatial dependencies, as well as their composite effects on the average AoI.
The resultant expression is, nevertheless, a bit involved. 
By assuming the positions of interference sensors being i.i.d. and constitutes a Poisson marked process \cite{2014On}, we can obtain a tight approximation to it.

\begin{corollary} \label{cor:cell}
\textit{Under the setting of a cellular network with power control, the average AoI over the typical link under FSA updating protocol can be approximated by the following: 
\begin{align}
    &\bar{\Delta}_0^\mathrm{C} 
    \nonumber\\
    &\approx \frac{F^2\!\!-\!\!1}{12F}\eta \int_0^\infty \!\! e^{-z-C_1 z^{1-\epsilon}}\! \mathrm{d}z \! + \! \frac{F}{\eta} \int^\infty \!\! e^{-z+C_2 z^{1-\epsilon}}\! \mathrm{d}z \! + \! \frac{1\!-\!F}{2}
    \nonumber\\
    &= \frac{1\!-\!F}{2} \!+\! \sum_{n=0}^\infty \frac{\Gamma \big(1\!+\!(1\!-\! \epsilon)n \big)}{n!} \bigg(\frac{F^2\!\!-\!\!1}{12F}\eta (-C_1)^n \!+\! \frac{F}{\eta} {C_2}^n \bigg),
\end{align}
where $C_1=\frac{\lambda_\mathrm{s}}{\lambda_\mathrm{d}} \frac{\eta}{F} \theta^\delta \Gamma(1-\delta) \Gamma(1+\delta) \Gamma(1+\epsilon)$ and $C_2=\frac{\lambda_\mathrm{s}}{\lambda_\mathrm{d}} \frac{\eta}{F} \big(1-\frac{\eta}{F} \big)\theta^\delta \Gamma(1-\delta) \Gamma(1+\delta) \Gamma(1+\epsilon)$.}
\begin{IEEEproof}
Please see Appendix~G.
\end{IEEEproof}   
\end{corollary}

In order to garner more insights from Theorem~\ref{thm:Cellular_AveAoI}, we resort to the following special cases. 
\subsubsection{When $F=1$}
The average AoI given in \eqref{equ:cell_ave} reduces to that under SA protocol, which can be expressed as:
\begin{align} \label{equ:SA_ave}
    \bar{\Delta}_0^\textit{SA} &=\mathbb{E} \left[\frac{1}{\eta \mu_0^\Phi}\right]
    \nonumber\\
    &=\frac{1}{\eta} \int_0^\infty \!\! \exp{\left(\! - z +\frac{\lambda_\mathrm{s}}{\lambda_\mathrm{d}} \eta g_\theta(1 \!- \!\eta,z;\alpha,\epsilon) \right)}\mathrm{d}z.
\end{align}
This result accounts for the effect of power control on the AoI performance in a Poisson cellular network,  which complements the current development of AoI analysis in wireless networks. 

\subsubsection{When $F>1$}
Similar to the scenario under the Poisson bipolar network, we can rewrite \eqref{equ:cell_ave} 
as follows:
\begin{align}\label{equ:CelAveAoI_EffctUpdtRt}
    &\bar{\Delta}_0^\mathrm{C} = \bar{\Delta}_0^\textit{SA}(\beta)
    \nonumber\\
    & + \underbrace{\frac{F^2 - 1}{12}\beta \!\!\int_0^\infty \!\!\!\!\! \exp{\Big(-z- \frac{\lambda_\mathrm{s}}{\lambda_\mathrm{d}} \beta g_\theta(1,z;\alpha,\epsilon) \Big)}\mathrm{d}z \!+\! \frac{1\! -\! F}{2} }_{Q_2},
\end{align}
where $\bar{\Delta}_0^\textit{SA}(\beta)$ denotes the average AoI under the SA update protocol when the update rate is $\beta$.
From \eqref{equ:CelAveAoI_EffctUpdtRt} we can see that FSA is not effectual in reducing average AoI unless $Q_2$ is negative.
Such a condition can be formally rewritten as the following:
\begin{align}
    \int_0^\infty \exp{\Big( -z - \frac{\lambda_\mathrm{s}}{2\lambda_\mathrm{d}} \eta g_\theta(1,z;\alpha,\epsilon) \Big)}\mathrm{d}z < \frac{4}{\eta}.
\end{align}

As power control plays a critical role in the AoI performance, we investigate the average AoI under three specific cases. 

\subsubsection{No power control}
In this case, $\epsilon=0$, each sensor transmits information with the same power.
Correspondingly, we can derive the average AoI over the typical link as follows:
 \begin{align}\label{equ:NoPC_AveAoI}
     \bar{\Delta}_0^\mathrm{C,N} \!\! = \!
     \begin{cases}
     \infty,  &\mbox{if } \mbox{$\frac{\lambda_s}{\lambda_d}\Omega_\delta \frac{\eta}{F} (1\!\!-\!\!\frac{\eta}{F})^{\delta-1} \geq 1$}, \\
     \frac{\frac{F}{\eta}}{1\!-\!\frac{\lambda_s}{\lambda_d}\Omega_\delta \frac{\eta}{F} (1\!-\!\frac{\eta}{F})^{\delta\!-\!1}} \!\!\!\!\!\!&+ \frac{\frac{F^2\!-\!1}{12F}\! \times \! \eta}{1 + \frac{\lambda_s}{\lambda_d}\frac{\eta}{F}\Omega_\delta}\!+\!\frac{1\!-\!F}{2},  \mbox{ otherwise},
     \end{cases}
 \end{align}
where $\Omega_\delta = \theta^\delta \Gamma(1-\delta) \Gamma(1+\delta)$.
The result in \eqref{equ:NoPC_AveAoI} provides a closed-form expression that explicitly illustrates the influence of spatial contention on the average AoI, from which we can see that increasing the density of sensors degrades the average AoI reciprocally, as they need to vie for the ratio resources to transmit information packets.

\subsubsection{Full path inversion}
In this case, $\epsilon=1$, and the impact of path loss on the useful signal power can be alleviated by power control.
Nonetheless, interference may also go up as other transmitters are also raising their transmit power.
Consequently, we can derive the average AoI over the typical link under FSA protocol as follows:
\begin{align}
    &\bar{\Delta}_0^\mathrm{C,F} \!\!=\! \frac{F^2\!\!-\!\!1}{12F}\eta \cdot \exp\! {\Bigg( \frac{4}{5} \frac{\lambda_\mathrm{s}}{\lambda_\mathrm{d}} \frac{\eta}{F} \int_0^\infty \!\!\! \int_0^1 \! \frac{ze^{-zs} \mathrm{d}s \mathrm{d}z }{(1+\frac{s^\delta}{\theta})(1\!-\!e^{-z})} \Bigg)}
    \nonumber\\
    &+\!\frac{F}{\eta} \exp\!{\Bigg(\frac{4}{5} \frac{\lambda_\mathrm{s}}{\lambda_\mathrm{d}} \frac{\eta}{F} \! \int_0^\infty \!\!\! \int_0^1 \! \frac{ze^{-zs} \mathrm{d}s \mathrm{d}z }{(1\!\!-\!\!\frac{\eta}{F}\!+\!\frac{s^\delta}{\theta})(1\!-\!e^{-z})} \Bigg)} \!+\frac{1\! -\!F}{2}.
\end{align}

Additionally, in view of the accurate expression being complicated, we adopt Corollary~\ref{cor:cell} to get an approximated result as follows:
\begin{align}
    \bar{\Delta}_0^\mathrm{C,F} &\approx \frac{F^2\!-\!1}{12F}\eta \exp\left( - \frac{ 2 \eta \lambda_{ \mathrm{s} } }{ F \lambda_{ \mathrm{d} } } \Omega_\delta \right) 
    \nonumber\\
    &+\frac{F}{\eta} \exp\left( \frac{ 2 \eta \lambda_{ \mathrm{s} } }{ F \lambda_{ \mathrm{d} } } \Omega_\delta \Big( 1 - \frac{\eta}{F} \Big)^{ \delta - 1 } \right) + \frac{1\! -\!F}{2}.
\end{align}

\subsubsection{Maximum power constraint}
In practice, transmission power is often limited to a maximum value. 
Therefore, in this part, we consider a maximum power constraint model, which is described as follows:
\begin{align}
  P_i = 
  \begin{cases}
  P_\mathrm{tx}R_i^{\alpha \epsilon}, &R_i<p^{\frac{1}{\alpha \epsilon}}, \\
  P_\mathrm{max}, &R_i \geq p^{\frac{1}{\alpha \epsilon}},
  \end{cases}
\end{align}
where $P_\mathrm{max}$ indicates the maximum transmission power, and $p=\frac{P_\mathrm{max}}{P_\mathrm{tx}}$.
In this case, we can derive the average AoI over the typical link as follows:
\begin{align}
    &\bar{\Delta}_0^\mathrm{C,M} = \frac{1\!-\!F}{2}+
    \nonumber\\
    & \frac{F^2\!\!-\!\!1}{12} \eta \!\Bigg[\!\!\int_0^{C_\epsilon} \!\!\!\!\! \exp \! \Big(\!\!-\!\!z\!+\!g_\theta^{C_\epsilon}\!((1,\!0),\!z;\!\alpha,\!\epsilon)\!+\!g_{\theta,1}^{C_\epsilon}\!((1,\!\frac{\epsilon}{\delta},\!1),\!z;\!\alpha,\!\epsilon) \!\Big)\!\mathrm{d}\!z 
    \nonumber \\
    &+\!\! \int_{C_\epsilon}^\infty \!\!\! \exp \! \Big(\!\!-\!\!z\!+\!g_\theta^{C_\epsilon}\!((1,\!\frac{\epsilon}{\delta}),\!z;\!\alpha,\!\epsilon) \!+\! g_{\theta,\!-\!1}^{C_\epsilon}\!((1,\!0,\!1),\!z;\!\alpha,\!\epsilon)\!\Big)\!\mathrm{d}\!z \Bigg]
    \nonumber \\ 
    &+ \frac{F}{\eta}\!\Bigg[\!\!\int_0^{C_\epsilon} \!\!\!\!\! \exp \! \Big(\!\!-\!\!z\!+\!g_\theta^{C_\epsilon}\!((0,\!0),\!z;\!\alpha,\!\epsilon)\!+\!g_{\theta,1}^{C_\epsilon}\!((0,\!\frac{\epsilon}{\delta},\!-\!1),\!z;\!\alpha,\!\epsilon)\! \Big)\!\mathrm{d}\!z
    \nonumber \\
    &+\!\! \int_{C_\epsilon}^\infty \!\!\! \exp \!\Big(\!\!-\!\!z\!+\!g_\theta^{C_\epsilon}\!((0,\!\frac{\epsilon}{\delta}),\!z;\!\alpha,\!\epsilon)\!+\! g_{\theta,\!-\!1}^{C_\epsilon}\!((1,\!0,\!-\!1),\!z;\!\alpha,\!\epsilon) \!\Big)\!\mathrm{d}\!z\! \Bigg]\!,
\end{align}
where $C_\epsilon=\frac{5}{4}\lambda_\mathrm{d} \pi p^{\frac{\delta}{\epsilon}}$, while $g_\theta^{C_\epsilon}((m,\!a),\!z;\!\alpha,\!\epsilon)$ and $g_{\theta,n}^{C_\epsilon}((m,\!a,\!b),\!z;\!\alpha,\!\epsilon)$ are respectively given as
\begin{align}
    g_\theta^{C_\epsilon}((m,\!a),\!z;\!\alpha,\!\epsilon)\!=\! \frac{4}{5}z^2 \!\!\! \int_0^\infty \!\!\!\! \int_0^1 \! \frac{q e^{-zqs}(1\!-\!e^{-\frac{48}{25}zq}) \mathrm{d}s \mathrm{d}q}{(m \!+\! \frac{q^{\frac{\alpha(1-\epsilon)}{2}}}{\theta s^\frac{\alpha \epsilon}{2}}(\frac{C_\epsilon}{z})^a)(1\!-\!e^{-zq})},
\end{align}
\begin{align}
    &g_{\theta,n}^{C_\epsilon}((m,\!a,\!b),\!z;\!\alpha,\!\epsilon)\!=\! \frac{4}{5}z^2 \!\! \int_{C_\epsilon z^n}^\infty \!\!\! \int_{C_\epsilon z^n}^1 \! \frac{q e^{-zqs}(1\!-\!e^{-\frac{48}{25}zq}) }{1-e^{-zq}}
    \nonumber\\
    &\bigg[\frac{1}{\big(m\!+\!\theta q^{-\frac{\alpha}{2}}(\frac{C_\epsilon}{z})^a \big)^b} - \frac{1}{\big(m\!+\!\theta s^{\frac{\alpha \epsilon}{2}}q^{\frac{\alpha(\epsilon-1)}{2}}(\frac{C_\epsilon}{z})^{a-\frac{ \epsilon}{\delta}}\big)^b}\bigg]\mathrm{d}s \mathrm{d}q.
\end{align}

\setcounter{equation}{\value{equation}}
\setcounter{equation}{39}
\begin{figure*}[t!]
\begin{align} \label{equ:vari_cell}       
    \sigma^2_{ \Delta^\mathrm{C}_0 }&=\int_0^\infty e^{-z} \Bigg[ \frac{2F^2}{\eta^2}  \exp{\bigg( \frac{\lambda_\mathrm{s}}{\lambda_\mathrm{d}} \frac{2\eta}{F} G_\theta((1\!\!-\!\!\frac{\eta}{F},\!2,\!2,\!1\!\!-\!\!\frac{\eta}{F},\!1,\!\frac{\eta}{F}),z;\alpha,\epsilon) \bigg)}
    - \frac{F^2}{\eta^2}\exp{\bigg( \!-\!z\!+\! \frac{\lambda_\mathrm{s}}{\lambda_\mathrm{d}} \frac{2\eta}{F} G_\theta((1\!\!-\!\!\frac{\eta}{F},\!1,\!0,\!0,\!0,\!0),z;\alpha,\epsilon) \bigg)} 
    \nonumber\\
    &- \frac{(F^2\!\!-\!\!1)^2\eta^2}{144F^2}\exp{\bigg( \!-\!z\!-\! \frac{\lambda_\mathrm{s}}{\lambda_\mathrm{d}} \frac{2\eta}{F} G_\theta((1,\!1,\!0,\!0,\!0,\!0),z;\alpha,\epsilon) \bigg)}
    -\frac{F^2\!\!-\!\!1}{6}\exp{\bigg( \!-\!z\!+\! \frac{\lambda_\mathrm{s}}{\lambda_\mathrm{d}} (\frac{\eta}{F})^2 [G_\theta((1\!\!-\!\!\frac{\eta}{F},\!1,\!1,\!1,\!-1,\!0),z;\alpha,\epsilon) \bigg)}
    \nonumber\\
    &-\frac{F^2}{\eta}\exp{\bigg( \frac{\lambda_\mathrm{s}}{\lambda_\mathrm{d}} \frac{\eta}{F} G_\theta((1\!\!-\!\!\frac{\eta}{F},\!1,\!0,\!0,\!0,\!0),z;\alpha,\epsilon) \bigg)}
    +\frac{(F^2-1)\eta}{12}\exp{\bigg(  \frac{\lambda_\mathrm{s}}{\lambda_\mathrm{d}} \frac{\eta}{F} G_\theta((1,\!1,\!0,\!0,\!0,\!0),z;\alpha,\epsilon) \bigg)} \Bigg]\mathrm{d}z+\frac{F^2+2F-3}{4}
\end{align}
\setcounter{equation}{\value{equation}}{}
\setcounter{equation}{40}
\centering \rule[0pt]{18cm}{0.3pt}
\end{figure*}
\setcounter{equation}{40}

Similarly, we can also obtain the analytical expression for the variance of AoI by deconditioning $\mu_0^\Phi$ in \eqref{equ:varience_AoI} and taking the second moment of $\mu_0^\Phi$ minus the square of the average.
\begin{theorem} \label{the:bipo_var}
\textit{In a Poisson cellular network, the variance of AoI over the typical link under the FSA updating protocol can be expressed as \eqref{equ:vari_cell} at the top of this page, where 
\begin{align}
    &G_\theta((a,\!b,\!c,\!d,\!l,\!\varrho),\!z;\!\alpha,\!\epsilon) 
    \nonumber\\
    &= \frac{4}{5} z^2\!\! \int_0^\infty \!\!\!\! \int_0^1\!\! \frac{q \exp{(-zqs)} \mathrm{d}s \mathrm{d}q}{\frac{(a+\frac{q^{\alpha(1-\epsilon)/2}}{\theta s^{\alpha \epsilon/2}})^b}{c(d+\frac{q^{\alpha(1\!-\!\epsilon)/2}}{\theta s^{\alpha \epsilon/2}})^l+\varrho}(1-e^{-zq})}.
\end{align}
}
\end{theorem}
\begin{IEEEproof}
Similar to the derivation in \eqref{equ:cell_1}, we can calculate the second moment of $\frac{1}{\mu_0^\Phi}$ as follows:
\begin{align} \label{equ:cell_-2}
    &\mathbb{E}\left[\frac{1}{(\mu_0^\Phi)^2}\right] = \mathbb{E} \left[ \prod_{i \neq 0} \Bigg( 1+\frac{\eta/F}{1-\frac{\eta}{F}+\frac{D_i^\alpha}{\theta R_i^{\alpha \epsilon} r^{\alpha (1\!-\! \epsilon)}}} \Bigg)^2 \right]
    \nonumber\\
    &=\!\! \mathbb{E} \Bigg[\! \prod_{i \neq 0}\! \int_0^{D_i} \!\! \bigg(\! 1\!+\!\frac{\eta/F}{1\!\!-\!\!\frac{\eta}{F}\!\!+\!\!\frac{D_i^\alpha u^{\!-\!\alpha \epsilon}}{\theta r^{\alpha (1\!-\! \epsilon)}}}\! \bigg)^{\!2} \frac{\frac{5}{2}\lambda_\mathrm{d} \pi u \exp{(\!-\!\frac{5}{4} \lambda_\mathrm{d} \pi u^2)} \mathrm{d}u}{1\!-\!\exp{(\!-\!\frac{5}{4} \lambda_\mathrm{d} \pi D_i^2)}} \Bigg]
    \nonumber\\
    &=\!\! \mathbb{E} \Bigg[ \exp \!\bigg(2\pi \lambda_\mathrm{s}\!\!\int_0^\infty \!\!\!\! \int_0^\kappa \!\! \Big(2\!+\!\frac{\eta/F}{1\!-\!\frac{\eta}{F}\!+\!\frac{D_i^\alpha u^{-\alpha \epsilon}}{\theta r^{\alpha(1-\epsilon)}}} \Big)
    \nonumber\\
    & \qquad \qquad \qquad  \times \frac{\frac{\eta}{F} \!\!\cdot\!\! \frac{5}{2}\lambda_\mathrm{d}\pi u \exp{(-\frac{5}{4}\lambda_\mathrm{d} \pi u^2)} \mathrm{d}u \kappa \mathrm{d}\kappa}{(1\!-\! \frac{\eta}{F}\!+\!\frac{\kappa^\alpha u^{-\alpha \epsilon}}{\theta r^{\alpha(1-\epsilon)}})(1\!-\!\exp{(-\frac{5}{4}\lambda_\mathrm{d} \pi \kappa^2)})} \bigg)\! \Bigg]
    \nonumber\\
    &= \!\!\! \int_0^\infty \!\!\! \exp{\! \Bigg( \!\!\! -\!\! z\!\!+\!\!\frac{4}{5}\frac{\lambda_\mathrm{s}}{\lambda_\mathrm{d}} \!z^2 \!\!\! \int_0^\infty \!\!\!\!\! \int_0^1 \!\! \frac{ \frac{\eta}{F}q e^{\!-\!zsq} \big( 2\!\!-\!\! \frac{\eta}{F} \!\!+\!\! \frac{2q^{\frac{\alpha(1\!-\! \epsilon)}{2}}}{\theta s^{\frac{\alpha \epsilon}{2}}} \big) \mathrm{d}s \mathrm{d}q}{\big( 1\!\!-\!\!\frac{\eta}{F}\!+\!\frac{q^{\frac{\alpha(1\!-\!\epsilon)}{2}}}{\theta s^{\frac{\alpha \epsilon}{2}}} \big)^2 \big( 1\!-\!e^{-zq} \big)} \!\Bigg)}\! \mathrm{d}z.
\end{align}
Substituting the expressions for the first moment of $\mu_0^\Phi$ as per \eqref{equ:cell_1}, while the first and the second moment of $\frac{1}{\mu_0^\Phi}$ given in \eqref{equ:cell_-1} and \eqref{equ:cell_-2} respectively, into the expression for the variance of AoI after deconditioning $\mu_0^\Phi$ \eqref{equ:bipo_sig}, we can obtain the result in this theorem.
\end{IEEEproof}

When $F=1$, this result also reduces to the variance of AoI under SA protocol, given as the following. 

\begin{corollary}
\textit{In a Poisson cellular network, the variance of AoI over the typical link SA protocol is given by
\begin{align}
    \sigma^2_{ \Delta^\textit{SA}_0 } &= \!\!\int_0^\infty \!\! e^{-z} \Bigg[ \frac{2}{\eta^2}  \exp{\bigg( 2\frac{\lambda_\mathrm{s}}{\lambda_\mathrm{d}} \eta g_\theta((1\!\!-\!\!\eta,\!2,\!2,\!1\!\!-\!\!\eta,\!1,\!\eta),z;\alpha,\epsilon) \bigg)}
    \nonumber\\
    &- \frac{1}{\eta^2}\exp{\bigg( \!-\!z\!+\! 2\frac{\lambda_\mathrm{s}}{\lambda_\mathrm{d}} \eta g_\theta((1\!\!-\!\!\eta,\!1,\!0,\!0,\!0,\!0),z;\alpha,\epsilon) \bigg)} 
    \nonumber\\
    &-\frac{1}{\eta}\exp{\bigg( \frac{\lambda_\mathrm{s}}{\lambda_\mathrm{d}} \eta g_\theta((1\!\!-\!\!\eta,\!1,\!0,\!0,\!0,\!0),z;\alpha,\epsilon) \bigg)}
    \Bigg]\mathrm{d}z.
\end{align}}
\end{corollary}

\begin{remark}
\textit{Given the specific values of the parameters, integrals given above can be numerically evaluated via using two integral functions, namely \textbf{integral} and \textbf{integral2}, in Matlab. 
The infinite upper limits can be expressed by \textbf{inf}
or replaced by a large and appropriate finite value according to the required accuracy.}
\end{remark}

\subsection{Numerical Results}
In this section, we utilize the developed theoretical expressions to numerically evaluate and investigate the effect of FSA on the average AoI in a cellular network. Unless otherwise specified, we use the same set of parameters as in Section \ref{sec:bipo_num}.

\begin{figure}[t!]
  \centering
  \subfigure[\label{fig:8a}]{\includegraphics[width=0.98\columnwidth]{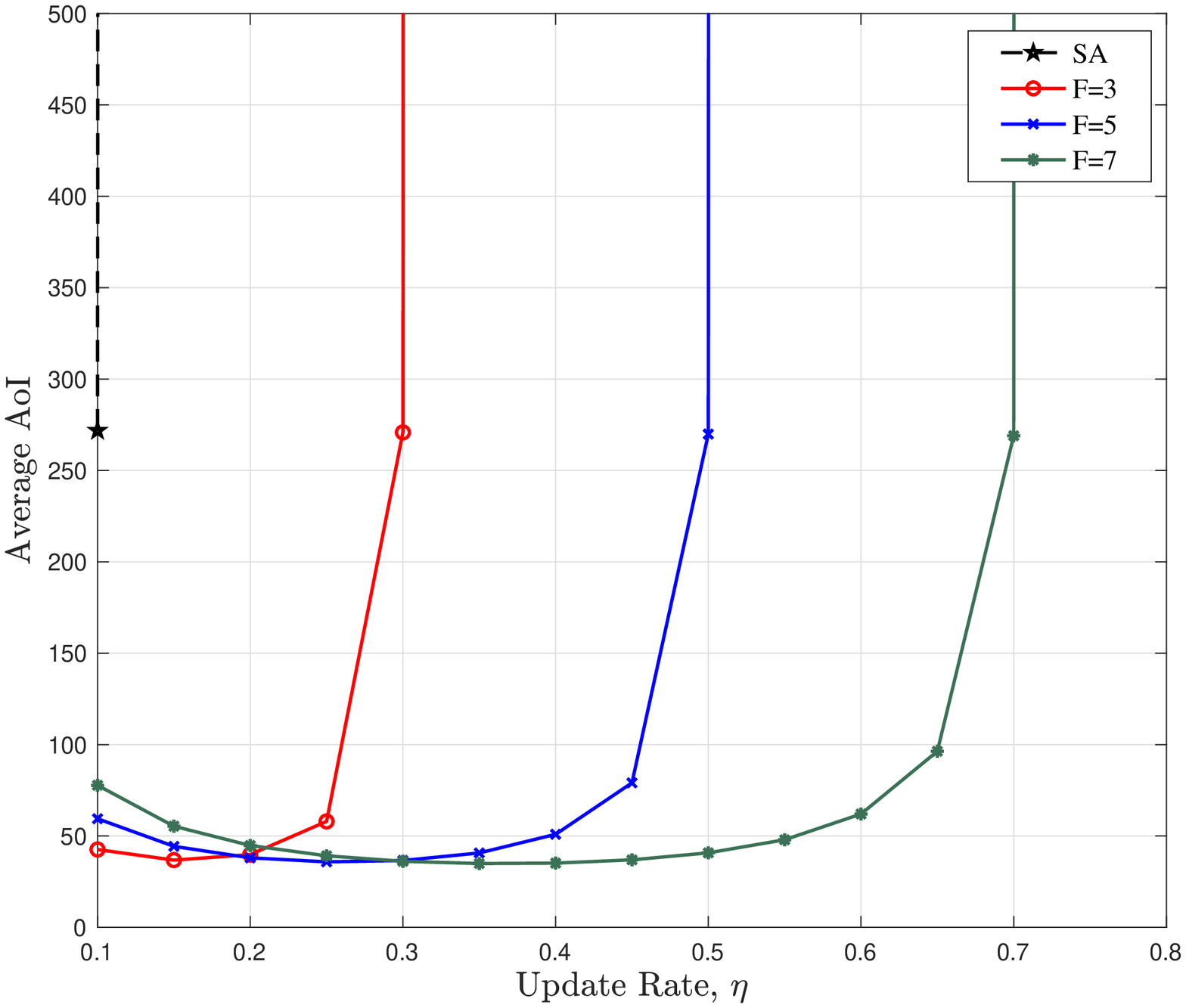}} ~
  \subfigure[\label{fig:8b}]{\includegraphics[width=0.98\columnwidth]{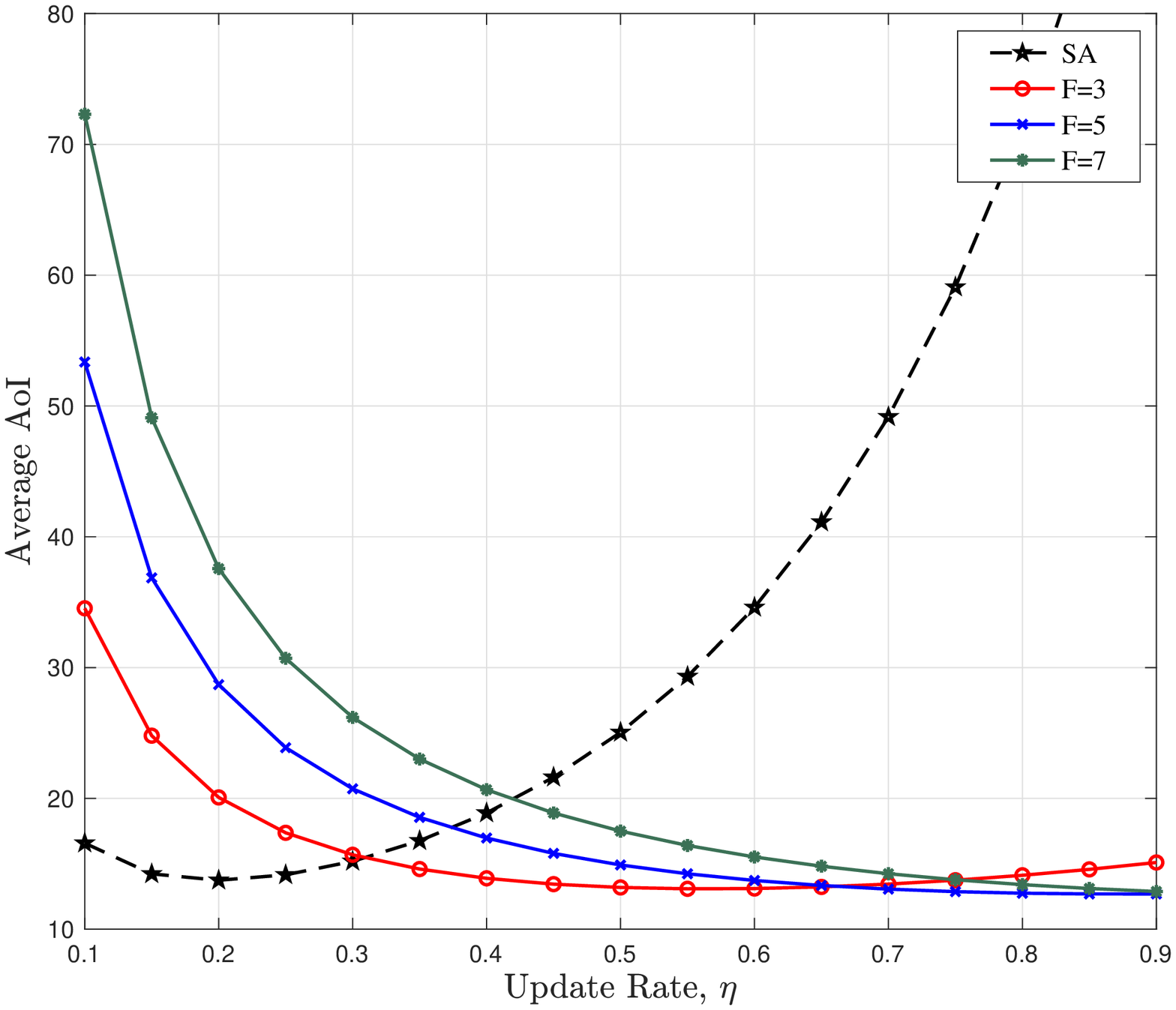}}
  \caption{Average AoI versus update rate under different framesizes $F$ and different power control strategies: ($a$) no power control ($\epsilon=0$) and ($b$) full path inversion ($\epsilon=1$), in which we set $\frac{\lambda_\mathrm{s}}{\lambda_\mathrm{d}} = 5$, and vary the framesize as $F=1$ (i.e. slotted ALOHA), $3$, $5$, and $7$.  }  
  \label{fig:cell_eta}
\end{figure}
Fig. \ref{fig:cell_eta} illustrates the relationship between the average AoI and update rate under different framesizes. In this figure, we compare the AoI performance under two different power control policies, namely, the unified transmit power and full path inversion. 
The figure reveals a similar phenomenon as that in the bipolar network, i.e., there exists an optimal $\eta$ that minimizes the average AoI, and as $\eta$ increases, the FSA updating protocol outperforms the traditional SA protocol.
Such an observation validates the importance of adopting frame structure in the status updating protocol. 
Besides, by comparing Fig.~\ref{fig:8a} and Fig.~\ref{fig:8b}, we see that using power control strategy can substantially reduce the average AoI, especially when there is no frame and/or $\eta$ is large.
Specifically, in Fig. \ref{fig:8a}, the average AoI under SA protocol increases sharply when $\eta$ is still small. 
As $F$ increases, AoI can maintain a lower value in a wider range of $\eta$. 
In Fig. \ref{fig:8b}, after the implementation of power control, the change of the average AoI becomes slower, especially the performance of SA is obviously better.

\begin{figure}[t!]
  \centering
  \subfigure[\label{fig:9a}]{\includegraphics[width=0.98\columnwidth]{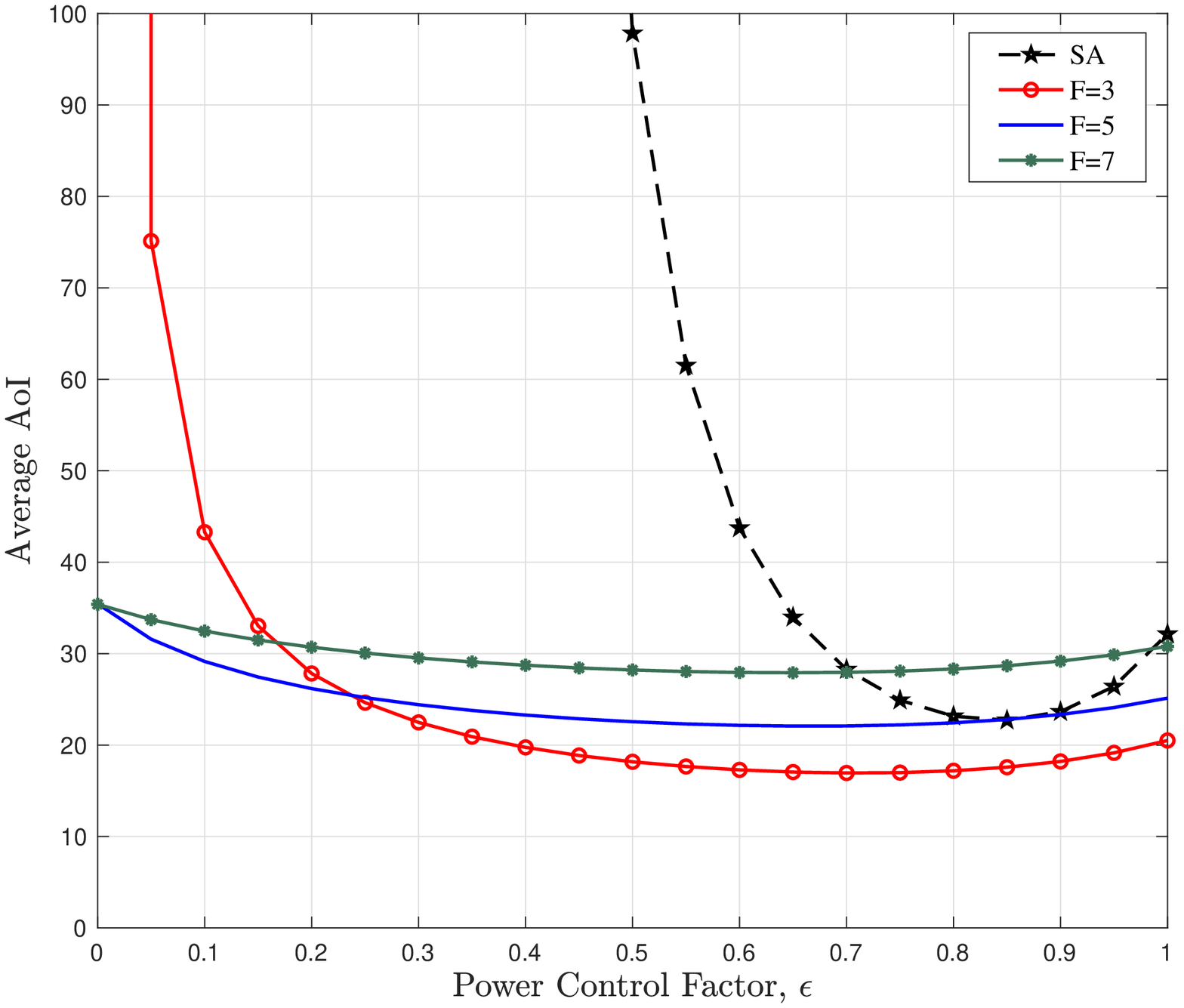}} ~
  \subfigure[\label{fig:9b}]{\includegraphics[width=0.98\columnwidth]{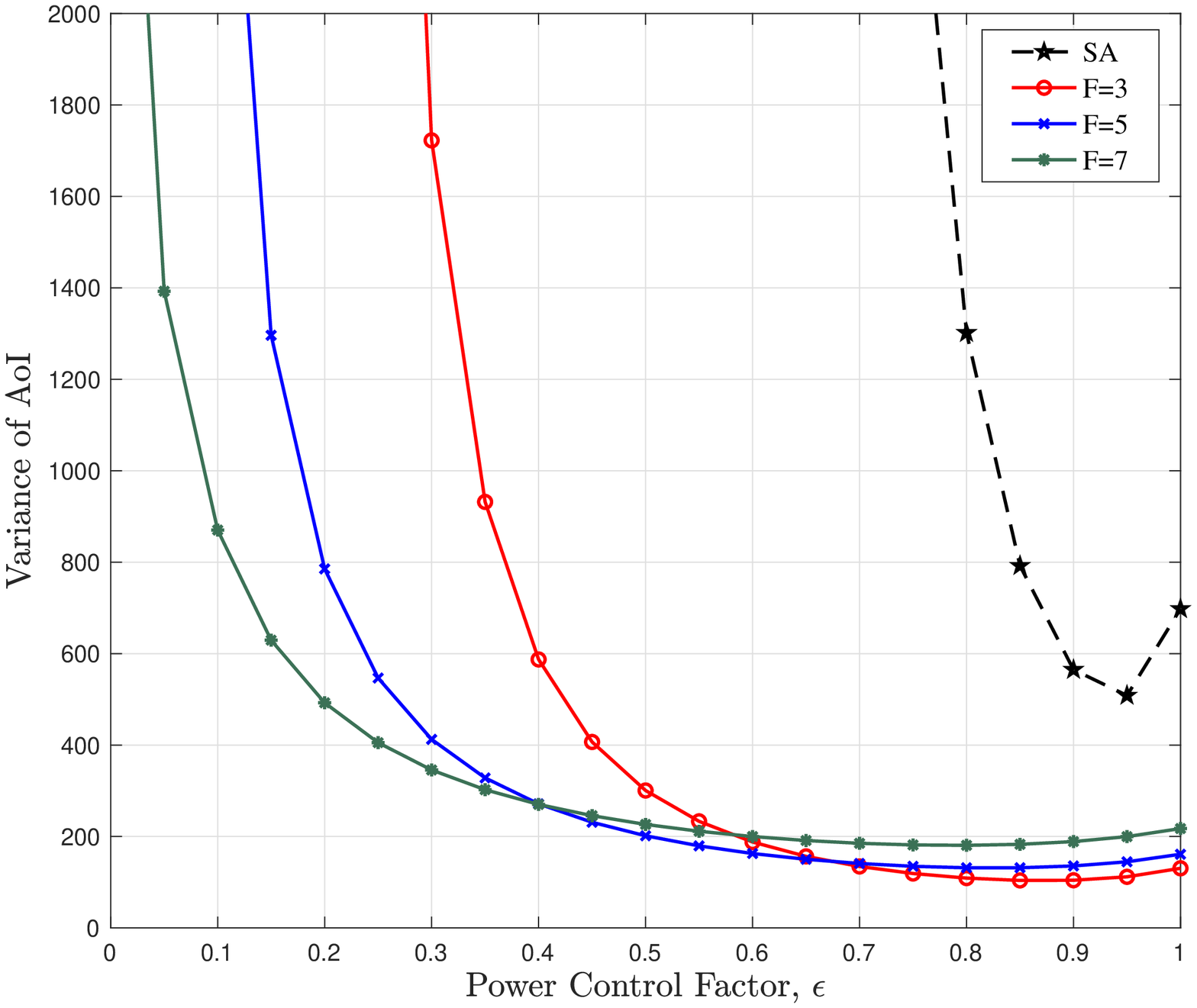}}
  \caption{AoI performance verse the power control factor $\epsilon$: ($a$) average AoI and ($b$) variance of AoI, in which we set $\eta=0.4$,  $\frac{\lambda_\mathrm{s}}{\lambda_\mathrm{d}} = 5$, and vary the framesize as $F=1$ (i.e. slotted ALOHA), $3$, $5$, and $7$.  }  
  \label{fig:pow_ctl}
\end{figure}
In Fig. \ref{fig:pow_ctl}, we plot the average and variance of AoI as a function of power control factor $\epsilon$ under different framesizes.
From this figure, we immediately notice that power control has a direct influence on the performance of AoI, whereas for a varying value of $F$, there exists an optimal $\epsilon$ that minimizes the average or variance of AoI. 
Moreover, the optimal $\epsilon$ is closely related to the particular value of $F$. 
To be more precise, when $F=1$, i.e., the updating protocol is SA, there are no frame to regularize the potential collisions amongst the transmitters. 
Since the typical source node could locate in a long distance to the data fussing center, the power attenuation caused by path loss, together with the deterioration from interference, can lead to a significant degradation to the average and variance of AoI. 
Therefore, one shall increase the power control factor to compensate the path loss. 
Nonetheless, increasing the source nodes' transmit power not only strengthens their signal power, especially for those located remotely to the receivers, but also enlarges the accumulated interference.
Hence, the power control factor needs to be adequately tuned so as to optimize the age performance. 
In contrast, under the FSA protocol (with $F \geq 2$), the adoption of frame reduces the competition between nodes and reducing the requirement of power control. As such, the optimal $\epsilon$ in FSA will be smaller than that under SA.
Additionally, we note that although a larger framesize provides more possibility of small interference transmission environment for remote sources, improves the probability of successful transmission, so that the network still maintains well performance, merely increasing $F$ does not always benefit the AoI. 

\begin{figure}[t!]
  \centering
  \subfigure[\label{fig:10a}]{\includegraphics[width=0.98\columnwidth]{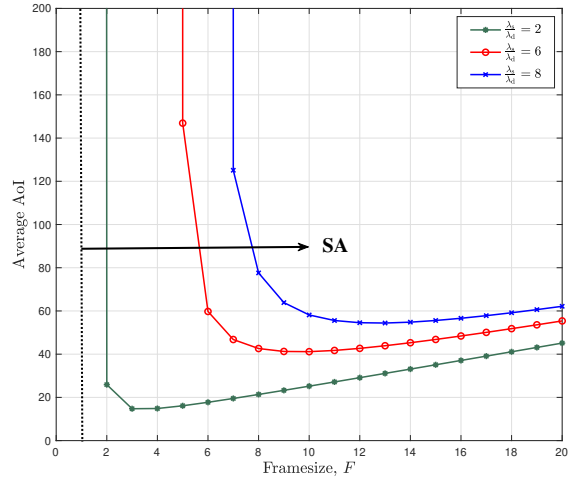}} ~
  \subfigure[\label{fig:10b}]{\includegraphics[width=0.98\columnwidth]{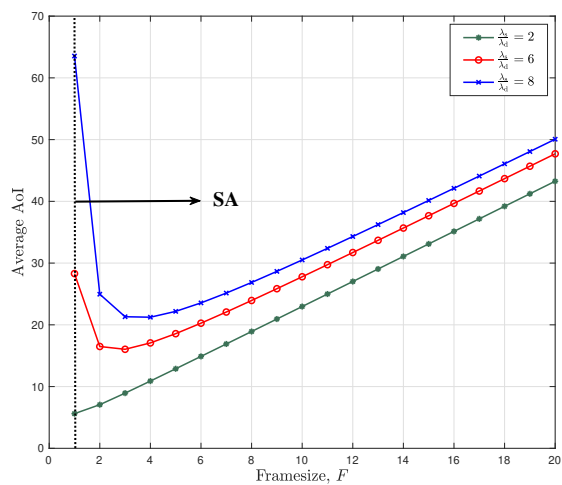}}
  \caption{ AoI performance versus framesize with different deployment densities $\frac{\lambda_\mathrm{s}}{\lambda_\mathrm{d}}$ and different power control strategies: ($a$) no power control ($\epsilon=0$) and ($b$) full path inversion ($\epsilon=1$), in which we set $\eta = 0.4$, and vary the average number of sources that each data fusing center can contact as $\frac{\lambda_\mathrm{s}}{\lambda_\mathrm{d}}=2$, $4$, $6$, and $8$.  }  
  \label{fig:cell_F}
\end{figure}
In Fig. \ref{fig:cell_F}, we depict the network average AoI as a function of $F$, under different deployment densities. 
Unlike the bipolar network, cellular network deployment affects AoI only in terms of the density ratio of sources and destination nodes $\frac{\lambda_\mathrm{s}}{\lambda_\mathrm{d}}$, that is, the average number of sensors connected to each data fusing center.
First, we observe a phenomenon similar to Fig. \ref{fig:bipo_F}, that is, depending on the network spacial deployment, there may be an optimal $F$, and increasing $\frac{\lambda_\mathrm{s}}{\lambda_\mathrm{d}}$ will definitely deteriorate AoI performance, but employing FSA protocol can alleviate this deterioration.
Second, we observe that the evolution trend of the average AoI with $F$ is different under different power control strategies. 
Specifically, Fig. \ref{fig:10a} shows that without power control, the average AoI under SA protocol has reached a very large value even when $\frac{\lambda_\mathrm{s}}{\lambda_\mathrm{d}}$ is still very small (e.g., $\frac{\lambda_\mathrm{s}}{\lambda_\mathrm{d}}=2$). 
However, the implementation of FSA strategy can obtain a relatively good performance, and with the increase of the ratio, we need a larger frame to maintain the average AoI performance in a better state.
In Fig. \ref{fig:10b}, the power control strategy can increase the competitiveness of effective signals and then improve the performance of AoI.
In this case, when $\frac{\lambda_\mathrm{s}}{\lambda_\mathrm{d}}$ is relatively small, FSA does not play a significant role, and even worse than the performance of AoI under SA protocol. 
As the ratio increases, namely the competition between sources increases, FSA will show its advantage.

\section{Conclusion} \label{sec:conclusion}
In this paper, we undertook an analytical study toward understanding the effect of FSA-like status updating protocol on the AoI performance in wireless networks. We adopted a general model that accounts for channel fading, path loss, and interference. We derived closed-form expressions for the average and its variance in Poisson bipolar and cellular networks, respectively.
Based on the analysis, we identified the operating regime under which FSA is instrumental in reducing AoI. 
We also provided the optimal framesize that minimizes the average AoI for a given configuration of network parameters.
The numerical results confirmed that when interference is severe, i.e., if the network is densely deployed and/or the source nodes are updating status information aggressively, employing FSA in the transmission protocol can substantially reduce the average and variance of AoI. 
In contrast, such a scheme is ineffective for AoI reduction in a sparsely deployed network, as interference is mild in this scenario, and the nodes should update more frequently to attain a small AoI. 
Our numerical results also showed that implementing FSA jointly with power control can further benefit the age performance in a wireless system, while the particular values of framesize and power control factor need to be adequately tuned to achieve the optimal gain. 

The mathematical framework presented in this paper lays the foundation for analyzing large-scale wireless networks with frame-based traffic patterns. 
The model can be extended to study the effects of FSA-based status updating protocols on the age performance under non-linear cost functions.
Additionally, studying the optimal design of frame structure with a variable size adapted to each transmitter's local geometry is another promising future research that can be extended from the fixed frame size setting and results shown in this paper.

\begin{appendix}
\subsection{Proof of Theorem 1} \label{pro:The1}
We adopt a graphical method to calculate the average AoI. 
Specifically, we denote by $X$ the number of frames between two successfully received updates, $I$ the interval between two consecutive updates received, $A$ the area between the two successful updates as illustrated in Figure~1, and use the subscript $w$ to represent the value of the $w$-th corresponding variable (e.g. define $I_w$ as the time elapsed since the $(w-1)$-th acceptance to the update before a new update is received again).
Thus, the time average AoI can be expressed as follows:
\begin{align}
    \bar{\Delta}_0
    =\lim_{W \to \infty}\frac{\sum_{w=1}^W A_w}{\sum_{w=1}^W I_w}
    =\frac{\mathbb{E}[\sum_{i=1}^I i]}{\mathbb{E}[I]} = \frac{1}{2}+\frac{\mathbb{E}[I^2]}{\mathbb{E}[I]},
\end{align}
in which $I_w$ is given by:
\begin{align} \label{equ:I}
    I_w=FX_w+(N_w-N_{w-1}),
\end{align}
numbering time slots in the frame, $N_w \in \{1,2,...,F \}$ represents the index of the time slot that is an update received for the $w$ time (e.g., in Figure~1, $N_{w-1}=3$ and $N_w=2$).
Note that $X$ obeys a geometric distribution with parameter $\eta\mu_0^\Phi$, therefore, $\mathbb{E}[X] = \frac{1}{\eta\mu_0^\Phi}$; $N_{w-1}$ and $N_{w}$ have the same distribution, being a discrete uniform distribution independent of $X$ in the range of $\{1,...F\}$. By denoting $\mathbb{E}[N]=\mathbb{E}[N_w]$, we have:
\begin{align} 
    \mathbb{E}[I]&=F\mathbb{E}[X],
    \\
    \mathbb{E}[I^2]&=F^2\mathbb{E}[X]+2(\mathbb{E}[N^2]-\mathbb{E}[N]^2).
\end{align}
On the other hand, $\mathbb{E}[N]$ and $\mathbb{E}[N^2]$ can be respectively computed as follows:
\begin{align}
    \mathbb{E}[N] &= \sum_{n=1}^F n \mathbb{P}[N=n] = \frac{1}{F} \sum_{d=1}^F d = \frac{F+1}{2},
    \\
    \mathbb{E}[N^2] &= \sum_{n=1}^F n^2 \mathbb{P}[N=n] = \frac{1}{F} \sum_{d=1}^F d^2 = \frac{(F+1)(2F+1)}{6}.
\end{align}
Combining the above fragments, we can obtain the conditional average AoI.

\subsection{Proof of Theorem 2} \label{pro:The2}
The similar graphical method is applied to Theorem~2, we can express the average quadratic AoI as follows:
\begin{align}
    \bar{\Delta}_0^2
    =\frac{\mathbb{E}[\sum_{i=1}^I i^2]}{\mathbb{E}[I]} = \frac{2\mathbb{E}[I^3]+3\mathbb{E}[I^2]+\mathbb{E}[I]}{6\mathbb{E}[I]}.
\end{align}
According to the expression for $I_w$ in \eqref{equ:I}, we can obtain:
\begin{align}
    \mathbb{E}[I^3]=F^3\mathbb{E}[X^3]+6F(\mathbb{E}[N^2]-\mathbb{E}[N]^2)\mathbb{E}[X].
\end{align}

Since $X$ obeys a geometric distribution with parameter $\eta \mu_0^\Phi$, we can carry out the following calculation:
\begin{align}
    \mathbb{E}[X^3] &= \sum_{k=1}^\infty k^3 \mathbb{P}(X=k)
    =\sum_{k=1}^\infty k^3 \eta \mu_0^\Phi (1-\eta \mu_0^\Phi)^{k-1}
    \nonumber \\
    &=-\eta \mu_0^\Phi \frac{\mathrm{d} \sum_{k=1}^\infty k^2 (1-\eta \mu_0^\Phi)^{k}}{\mathrm{d}(\eta \mu_0^\Phi)}
    \nonumber \\
    &=-\eta \mu_0^\Phi \frac{\mathrm{d} (1-\eta \mu_0^\Phi)\mathbb{E}[X^2]/(\eta \mu_0^\Phi)}{\mathrm{d}(\eta \mu_0^\Phi)}
    \nonumber \\
    &=\frac{(\eta \mu_0^\Phi)^2-6\eta \mu_0^\Phi+6}{(\eta \mu_0^\Phi)^3}.
\end{align}
Then, combining the above results, we get the result.

\subsection{Proof of Lemma 1} \label{pro:Lem1}
According to the transmission protocol, we note that every source decides whether it will update in a typical frame independently with probability $\eta$, and if a source decides to update in this frame, it randomly selects a time slot according to a uniform distribution.
Consequently, at any given time slot, a source activates with probability $\eta/F$, i.e., $\mathbb{P}(\nu_i=1)=\frac{\eta}{F}$.
As such, we can substitute \eqref{equ:SIR_PBN} into \eqref{equ:CndTXProb} and get the following:
\begin{align}
    \mu_0^\Phi &= \mathbb{P} \left(\frac{P_{ \mathrm{tx} }h_0r^{-\alpha}}{\sum_{i \neq 0}P_{ \mathrm{tx} }h_i\nu_i\parallel \!\! x_i \!\! \parallel ^{-\alpha}} >\theta \Big\vert \Phi \right)
    \nonumber\\
    &= \mathbb{P} \left(h_0 > \theta r^\alpha\sum_{i \neq 0}h_i \nu_i\parallel \!\! x_i \!\! \parallel ^{-\alpha}  \Big\vert \Phi \right)
     \nonumber\\
     &= \mathbb{E} \left[ \prod_{i \neq 0}\exp\left(-\theta r^\alpha h_i \nu_i\parallel \!\! x_i \!\! \parallel ^{-\alpha}\right) \Big\vert \Phi \right]
    \nonumber\\
     &\stackrel{(a)}{=} \mathbb{E} \left[ \prod_{i \neq 0} \frac{1}{1+\theta r^\alpha \nu_i \parallel \!\! x_i \!\! \parallel ^{-\alpha}} \right]
    \nonumber\\
     &\stackrel{(b)}{=}  \prod_{i \neq 0}\left(\frac{\eta/F}{1+\theta r^\alpha \parallel \!\! x_i \!\! \parallel ^{-\alpha}}+1-\frac{\eta}{F} \right),
\end{align}
where (a) follows since $\{ h_i \}_{i=1}^\infty$ are i.i.d. random variables following the exponential distribution with unit mean, and (b) follows as $\{ \nu_i \}_{i=1}^\infty$ are independent of each other.
Then, after a further simplification, we obtain the result shown in Lemma 1.

\subsection{Proof of Theorem 3} \label{pro:The3}
Deconditioning $\mu_0^\Phi$ in \eqref{equ:AoI}, we can get the expression of average AoI as:
\begin{align} \label{equ:aveAoI}
    \bar{\Delta}_0=\frac{F^2-1}{12F}\eta \mathbb{E}[\mu_0^\Phi] + \frac{F}{\eta} \mathbb{E}\bigg[\frac{1}{\mu_0^\Phi}\bigg]+\frac{1-F}{2}.
\end{align}
Using Lemma~\ref{lem:mu_0}, we can compute the mean of $\mu_0^\Phi$ as follows:
\begin{align} \label{equ:bipo_1}
    \mathbb{E}[\mu_0^\Phi] 
    &=\mathbb{E} \Bigg[\prod_{i \neq 0}\left(  1-\frac{\eta /F}{1+\Vert x_i \Vert ^{\alpha}/\theta r^\alpha} \right) \Bigg]
    \nonumber\\
    &= \mathbb{E} \Bigg[\exp \bigg(- \sum_{i \neq 0} - \log \bigg(1-\frac{\eta /F}{1+\Vert x_i \Vert ^{\alpha}/\theta r^\alpha} \bigg) \Bigg) \Bigg]
    \nonumber\\
    &\stackrel{(a)}{=} \exp \Bigg(\!-\!\lambda \int_{x\in \mathbb{R}^2} \left[ 1-\left( 1-\frac{\eta /F}{1+\Vert x \Vert ^\alpha/\theta r^\alpha} \right) \right] \mathrm{d}x \Bigg)
    \nonumber\\
    &\stackrel{(b)}{=} \exp \Bigg(- \lambda \pi \delta \int_0^{\infty} \frac{\eta/F}{1+u/\theta r^\alpha} u^{\delta-1} \mathrm{d}u \Bigg)
    \nonumber\\
    &\stackrel{(c)}{=} \exp \Bigg( -\lambda \pi r^2 \theta^\delta \Gamma(1-\delta) \Gamma(1+\delta) \frac{\eta}{F} \Bigg),
\end{align}
where (a) follows by using the probability generating functional (PGFL) of PPP \cite{2022Spatio}, (b) changes variables from rectangular to polar coordinates and sets $u= \Vert x \Vert ^\alpha$, and (c) is due to the result $\int_0^{\infty} \frac{u^{\delta-1}\mathrm{d}u}{u+m}=m^{\delta-1}\frac{\pi}{\sin{(\pi \delta)}}$ \cite{2015The}.
Similarly, we can calculate $\mathbb{E}[\frac{1}{\mu_0^\Phi}]$ as follows:
\begin{align} \label{equ:bipo_2}
    \mathbb{E}\bigg[\frac{1}{\mu_0^\Phi}\bigg] 
    &=\mathbb{E} \Bigg[\prod_{i \neq 0}\left(  1-\frac{\eta /F}{1+\Vert x_i \Vert ^{\alpha}/\theta r^\alpha} \right)^{-1} \Bigg]
    \nonumber\\
    &= \exp \Bigg(\!\!-\lambda \!\!\int_{x\in \mathbb{R}^2}\!\! \left[ 1-\left( 1 - \frac{\eta /F}{1+\Vert x \Vert ^\alpha/\theta r^\alpha} \right)^{\!\!-1} \right] \mathrm{d}x \Bigg)
    \nonumber\\
    &= \exp \Bigg(\!\!-\lambda \!\!\int_{x\in \mathbb{R}^2} \frac{\eta/F}{1-\frac{\eta}{F} + \frac{\Vert x \Vert ^\alpha}{\theta r^\alpha}} \mathrm{d}x \Bigg)
    \nonumber\\
    &= \exp \Bigg( \lambda \pi \delta \int_0^{\infty} \frac{\eta/F}{1-\frac{\eta}{F}+\frac{u}{\theta r^\alpha}} u^{\delta-1} \mathrm{d}u \Bigg)
    \nonumber\\
    &= \exp \Bigg( \lambda \pi r^2 \theta^\delta \Gamma(1-\delta) \Gamma(1+\delta) \frac{\eta}{F} \Big(1\!\!-\!\!\frac{\eta}{F} \Big)^{\delta-1} \Bigg).
\end{align}
By substituting \eqref{equ:bipo_1} and \eqref{equ:bipo_2} into \eqref{equ:aveAoI}, we can obtain the result shown in Theorem 1.

\subsection{Proof of Theorem 4} \label{pro:The4}
Similar to the steps taken in Appendix~\ref{pro:The3}, we can calculate the variance of AoI as follows:
\begin{align} \label{equ:bipo_sig}
    &\sigma^2_{ \Delta_0 } = \mathbb{E}[\Delta_0^2]-(\mathbb{E}[\Delta_0])^2
    \nonumber\\
    &=\! \frac{2F^2}{\eta^2}\mathbb{E}\!\left[ \frac{1}{(\mu_0^\Phi)^2} \right] \!-\! \frac{2F^2\!\!-\!\!F}{\eta}\mathbb{E}\!\left[ \frac{1}{\mu_0^\Phi} \right] \!+\! \frac{(F^2\!\!-\!\!1)\eta}{12F}  \mathbb{E}[\mu_0^\Phi] \!+\! \frac{F^2\!\!-\!\!F}{2}
    \nonumber\\
    &-\Bigg[ \frac{F^2-1}{12F}\eta \mathbb{E}[\mu_0^\Phi] + \frac{F}{\eta} \mathbb{E}\Big[\frac{1}{\mu_0^\Phi} \Big]+\frac{1-F}{2} \Bigg]^2,
\end{align}
where $\mathbb{E}[\frac{1}{(\mu_0^\Phi)^2}]$ can be derived by the following:
\begin{align}
    &\mathbb{E}\left[\frac{1}{(\mu_0^\Phi)^2}\right] = \mathbb{E} \Bigg[ \prod_{i \neq 0}\left(  1-\frac{\eta /F}{1+\Vert x_i \Vert ^{\alpha}/\theta r^\alpha} \right)^{-2} \Bigg]
    \nonumber \\
    &=\exp{\bigg( -\lambda \pi \delta \int_0^{\infty} \Big[ 1- \Big(1-\frac{\frac{\eta}{F}\theta r^{\alpha}}{u+\theta r^{\alpha}} \Big)^{-2} \Big] u^{\delta-1} \mathrm{d}u \bigg)}
    \nonumber \\
    &\stackrel{(a)}{=} \exp{\bigg( -\lambda \pi \delta \int_0^{\infty} \Big[ 1- \sum_{k=0}^\infty (k+1) \Big(\frac{\frac{\eta}{F}\theta r^{\alpha}}{u+\theta r^{\alpha}} \Big)^k \Big] u^{\delta-1} \mathrm{d}u \bigg)}
    \nonumber \\
    &=\exp{\bigg( \lambda \pi \delta \sum_{k=1}^{\infty} (k+1) \Big(\frac{\eta}{F}\theta r^{\alpha} \Big)^k \int_0^{\infty} \frac{ u^{\delta-1} \mathrm{d}u }{(u+\theta r^{\alpha})^k} \bigg)}
    \nonumber \\
    &\stackrel{(b)}{=}\exp{\bigg(- \lambda \pi \theta^\delta r^2\frac{\pi \delta}{\sin{(\pi \delta)}} \sum_{k=1}^{\infty}(k+1)\Big(-\frac{\eta}{F} \Big)^k \frac{\Gamma(\delta)}{\Gamma(k)\Gamma(\delta\!-\!k\!+\!1)} \bigg)}
    \nonumber \\
    &\stackrel{(c)}{=} \exp{\bigg( -C\sum_{k=1}^\infty (k+1) {\delta-1 \choose k-1}\Big(-\frac{\eta}{F} \Big)^k \bigg)},
\end{align}
where (a) leverages the relationship that $\frac{1}{(1+x)^2}=\sum_{k=0}^\infty(k+1)(-x)^k,\ \mathrm{for} \ \left| x \right|<1$, 
(b) uses 
\begin{align}
     \int_0^{\infty} \frac{ u^{\delta-1} \mathrm{d}u }{(u+\theta r^{\alpha})^k} = (\theta r^\alpha)^{\delta-k} \!\! \times \!\! \frac{(-1)^{k+1}\pi}{\sin(\pi \delta)} \!\!\times\!\! \frac{\Gamma(\delta)}{\Gamma(k)\Gamma(\delta\!-\!k\!+\!1)},
     \nonumber
\end{align}
and (c) follows from the Odd Element formula $\Gamma(\delta)\Gamma(1-\delta)=\frac{\pi}{\sin (\pi \delta)},  \ \mathrm{for}\ 0<\delta<1$ and the recursion of the Gamma function $\Gamma(1+\delta)= \delta \Gamma(\delta)$.
Then, by substituting the expression for $\mathbb{E}[\mu_0^\Phi]$, $\mathbb{E}[\frac{1}{\mu_0^\Phi}]$ and $\mathbb{E}[\frac{1}{(\mu_0^\Phi)^2}]$ into \eqref{equ:bipo_sig}, we get the result.

\subsection{Proof of Theorem 5} \label{pro:The5}
Since the sources only communicate with their nearest data fusing centers, when we condition on the distance between the user and the nearest center as $u$, then there is no data fusion center in area $A=2\pi u^2$.
Then, we can deduce that the distance $R$ between users and its dedicated receiver follows a distribution with the following probability density function:
\begin{align}
    f_R(u)=e^{-\lambda_\mathrm{d} \pi u^2}2\pi\lambda_\mathrm{d} u.
\end{align}

Similar to the derivation process of bipolar network, we can first obtain the conditional transmission success probability as:
\begin{align}
    \mu_0^\Phi = \prod_{i \neq 0}\left( 1-\frac{\eta/F}{1+\frac{D_i^\alpha}{\theta R_i^{\alpha \epsilon}r^{\alpha(1-\epsilon)}}} \right).
\end{align}
Next, we can calculate the first moment of $\mu_0^\Phi$ as follows:
\begin{align} \label{equ:cell_1}
    &\mathbb{E}[\mu_0^\Phi]=\mathbb{E} \Bigg[\! \prod_{i \neq 0} \!\left(\! 1\!\!-\!\!\int_0^{D_i} \!\!\! \frac{\frac{\eta}{F} \frac{5}{2}\lambda_\mathrm{d} \pi u \exp{(\!-\!\frac{5}{4}\lambda_\mathrm{d} \pi u^2)} \mathrm{d}u }{\big(1\!+\!\frac{D_i^\alpha u^{\!-\!\alpha \epsilon }}{\theta r^{\alpha (1\!-\!\epsilon)}} \big)\big(1\!-\!\exp{(\!-\!\frac{5}{4} \lambda_\mathrm{d}}\pi D_i^2)\big)}\!\! \right)\!\! \Bigg]
    \nonumber\\
    &= \mathbb{E} \Bigg[\! \exp \! \bigg(\! \sum_{i \neq 0} \! \log\Big(\! 1\!\!-\!\!\int_0^{D_i} \!\!\! \frac{\frac{\eta}{F} \frac{5}{2}\lambda_\mathrm{d} \pi u \exp{(\!-\!\frac{5}{4}\lambda_\mathrm{d} \pi u^2)} \mathrm{d}u }{\big(1\!+\!\frac{D_i^\alpha u^{\!-\!\alpha \epsilon }}{\theta r^{\alpha (1\!-\!\epsilon)}} \big) \! \big(1\!-\!e^{\!-\!\frac{5}{4} \lambda_\mathrm{d}\pi D_i^2}\big)}\! \Big)\! \bigg) \Bigg]
    \nonumber\\
    &\stackrel{(a)}{=}\mathbb{E}_r \Bigg[\! \exp{\Bigg(\!\! -\!2\pi \lambda_\mathrm{s} \int_0^\infty \!\!\!\! \int_0^\kappa \!\! \frac{\frac{\eta}{F} \frac{5}{2}\lambda_\mathrm{d} \pi u e^{\!-\!\frac{5}{4}\lambda_\mathrm{d} \pi u^2} \mathrm{d}u \kappa \mathrm{d}\kappa}{\big(1\!+\!\frac{\kappa^\alpha u^{\!-\!\alpha \epsilon }}{\theta r^{\alpha (1\!-\!\epsilon)}} \big)\big(1\!-\!e^{\!-\!\frac{5}{4} \lambda_\mathrm{d}\pi \kappa^2}\big)} \! \Bigg)\!} \Bigg]
    \nonumber\\
    &\stackrel{(b)}{=}\!\! \int_0^\infty \!\! \exp{\! \Bigg( \!\! -\!z\!-\!\frac{4}{5}\frac{\lambda_\mathrm{s}}{\lambda_\mathrm{d}}z^2 \!\! \int_0^\infty \!\!\!\! \int_0^1 \frac{\frac{\eta}{F}q\exp{(-zsq)} \mathrm{d}s \mathrm{d}q}{\big( 1\!+\!\frac{q^{\alpha(1\!-\!\epsilon)/2}}{\theta s^{\alpha \epsilon/2}} \big) \big( 1\!-\!e^{-zq} \big)} \Bigg)} \mathrm{d}z, 
\end{align}
where (a) follows from adopting the PGFL of PPP, and (b) results from the variable substitution: $s=\frac{u}{\kappa}$, $q=(\frac{\kappa}{r})^2$ and $z=\frac{5}{4}\lambda_\mathrm{d}\pi r^2$.
Similarly, we can calculate the expectation of $\frac{1}{\mu_0^\Phi}$ as follows:
\begin{align} \label{equ:cell_-1}
    &\mathbb{E}\bigg[\frac{1}{\mu_0^\Phi}\bigg] = \mathbb{E} \left[ \prod_{i \neq 0} \Bigg( 1+\frac{\eta/F}{1-\frac{\eta}{F}+\frac{D_i^\alpha}{\theta R_i^{\alpha \epsilon} r^{\alpha (1\!-\! \epsilon)}}} \Bigg) \right]
    \nonumber\\
    &=\!\! \int_0^\infty \!\! \exp{\! \Bigg( \!\! -\!z\!+\!\frac{4}{5}\frac{\lambda_\mathrm{s}}{\lambda_\mathrm{d}} \!\! \int_0^\infty \!\!\!\! \int_0^1 \!\! \frac{z^2 \frac{\eta}{F}q\exp{(-zsq)} \mathrm{d}s \mathrm{d}q}{\big( 1\!-\!\frac{\eta}{F}\!+\!\frac{q^{\alpha(1\!-\!\epsilon)/2}}{\theta s^{\alpha \epsilon/2}} \big) \big( 1\!-\!e^{-zq} \big)} \!\Bigg)} \mathrm{d}z.
\end{align}
Substituting \eqref{equ:cell_1} and \eqref{equ:cell_-1} into \eqref{equ:aveAoI}, we can obtain the average AoI of the cellular network.

\subsection{Proof of Corollary} \label{pro:Cor2}
We remove the spatial dependency of interference nodes' locations and approximate them as a marked PPP $\{x_i,W_{x_i}\}_{i=1}^\infty$, where $W_{x_i}=R_i^{\alpha \epsilon}$ \cite{2014On}.
Similar to the derivation process of Theorem~3, we can first condition on the distance of the typical transmitter-receiver pair $r$ and calculate the first moment of $\mu_0^\Phi$ as follows:
\begin{align} \label{equ:muPhi_r}
    &\mathbb{E}[\mu_0^\Phi \vert r] = \mathbb{E} \Bigg[\prod_{i \neq 0}\Bigg( 1-\frac{\eta/F}{1+\frac{D_i^\alpha}{\theta R_i^{\alpha \epsilon}r^{\alpha(1-\epsilon)}}} \Bigg) \Bigg]
    \nonumber\\
    &\approx \mathbb{E} \Bigg[\exp{\Bigg(\!-\!\lambda_\mathrm{s} \int_{x \in \mathbb{R}^2} \bigg[1-\bigg(1-\frac{\eta/F}{1+\frac{\Vert x \Vert^\alpha}{\theta W_x r^{\alpha(1-\epsilon)}}} \bigg) \bigg] \mathrm{d}x \Bigg)} \Bigg] 
    \nonumber \\
    &= \exp{ \Bigg(-\mathbb{E} \bigg[\int_{x \in \mathbb{R}^2} \frac{\lambda_\mathrm{s} \eta/F \mathrm{d}x }{1+\frac{\Vert x \Vert^\alpha}{\theta W_x r^{\alpha(1-\epsilon)}} }\bigg] \Bigg)}
    \nonumber \\
    &= \exp{ \Bigg(-\lambda_\mathrm{s} \frac{\eta}{F} \pi \mathbb{E} \bigg[\int_0^\infty \theta W_x r^{\alpha(1-\epsilon)} \frac{\delta u^{\delta-1} \mathrm{d}u}{u+\theta W_x r^{\alpha(1-\epsilon) }}\bigg] \Bigg)}
    \nonumber\\
    &= \exp{ \Bigg(-\lambda_\mathrm{s} \frac{\eta}{F} \pi \theta^\delta \Gamma(1-\delta) \Gamma(1+\delta) r^{2(1-\epsilon)}\mathbb{E}\big[ W_x^\delta \big] \Bigg)}
    \nonumber\\
    &\stackrel{(a)}{=} \exp{ \Bigg(-\lambda_\mathrm{s} \frac{\eta}{F} \pi \theta^\delta \Gamma(1-\delta) \Gamma(1+\delta) \frac{\Gamma(\epsilon+1)}{(\lambda_\mathrm{d}\pi)^{\epsilon}} r^{2(1-\epsilon)} \Bigg)},
\end{align}
where (a) results from the following
\begin{align}
    \mathbb{E}\big[ W_x^\delta \big] &= \mathbb{E}\big[R_i^{2 \epsilon} \big]
    \approx \int_0^\infty 2\pi \lambda_\mathrm{d}u e^{-\lambda_\mathrm{d} \pi u^2} u^{2 \epsilon} \mathrm{d}u 
    \nonumber\\
    &= \int_0^\infty \frac{t^\epsilon}{(\lambda_\mathrm{d} \pi)^\epsilon} e^{-t} \mathrm{d}t
    = \frac{\Gamma(\epsilon+1)}{(\lambda_\mathrm{d} \pi)^\epsilon}.
\end{align}
Then, we decondition \eqref{equ:muPhi_r} on $r$ and get the following:
\begin{align}
    &\mathbb{E}[\mu_0^\Phi] \approx \int_0^\infty 2\pi \lambda_\mathrm{d}r e^{-\lambda_\mathrm{d} \pi r^2} \mathbb{E}[\mu_0^\Phi \vert r] \mathrm{d}r 
    \nonumber\\
    & = \int_0^\infty \!\! \exp \Big(\!-\!z\!-\!\frac{\lambda_\mathrm{s}}{\lambda_\mathrm{d}} \frac{\eta}{F} \theta^\delta \Gamma(1\!-\!\delta) \Gamma(1\!+\!\delta) \Gamma(1\!+\!\epsilon) z^{1-\epsilon} \Big) \mathrm{d}z.
\end{align}
The expression above can be equivalently written as follows:
\begin{align}
    \mathbb{E}[\mu_0^\Phi] \approx \mathbb{E} \big[e^{-C_1 Z^{1-\epsilon}} \big]
\end{align}
where $Z$ is a random variable that obeys the exponential distribution with unit mean. 
Denote by $Y=Z^{1-\epsilon}$, we can compute the probability density function (pdf) of $Y$ as follows:
\begin{align}
    f_Y(y)&=\frac{\mathrm{d} \mathbb{P}(Y \leq y) }{\mathrm{d} y}
    =\frac{\mathrm{d} \mathbb{P} \big(Z \leq y^{\frac{1}{1-\epsilon}} \big) }{\mathrm{d} y}
    \nonumber\\
    &=\frac{\mathrm{d} \big(1-e^{-\frac{y}{1-\epsilon}} \big) }{\mathrm{d} y}
    = \frac{1}{1-\epsilon} y^{\frac{1}{1-\epsilon}-1} e^{-\frac{y}{1-\epsilon}}.
\end{align}
This expression indicates that $Y$ follows a Weibull distribution with the shape parameter $(1-\epsilon)$ and unit scale parameter. 
Therefore, we can obtain the approximate result as $\mathbb{E}[\mu_0^\Phi] \approx \mathbb{E} \big[e^{-C_1 Z^{1-\epsilon}} \big] = \sum_{n=0}^\infty \frac{(-C_1)^n}{n!}\Gamma \big(1+(1-\epsilon)n \big)$.

In the same way, we can calculate $\mathbb{E}\Big[\frac{1}{\mu_0^\Phi} \vert r \Big]$ as follows:
\begin{align}
    &\mathbb{E}\Big[\frac{1}{\mu_0^\Phi} \big\vert r \Big] = \mathbb{E} \Bigg[\prod_{i \neq 0}\Bigg( 1+\frac{\eta/F}{1-\frac{\eta}{F}+\frac{D_i^\alpha}{\theta R_i^{\alpha \epsilon}r^{\alpha(1-\epsilon)}}} \Bigg) \Bigg]
    \nonumber\\
    &\approx \exp{ \Bigg(\mathbb{E} \bigg[\int_{x \in \mathbb{R}^2} \frac{\lambda_\mathrm{s} \eta/F \mathrm{d}x }{1-\frac{\eta}{F}+\frac{\Vert x \Vert^\alpha}{\theta W_x r^{\alpha(1-\epsilon)}} }\bigg] \Bigg)}
    \nonumber \\
    &= \exp{\! \Bigg(\!\lambda_\mathrm{s} \frac{\eta}{F} \Big(1\!\!-\!\!\frac{\eta}{F} \Big)^{\delta\!-\!1}\! \pi \theta^\delta r^{2(1-\epsilon)} \Gamma(1\!-\!\delta) \Gamma(1\!+\!\delta) \frac{\Gamma(\epsilon\!+\!1)}{(\lambda_\mathrm{d}\pi)^{\epsilon}} \! \Bigg)}.
\end{align}
By deconditioning the above with respect to $r$, we can obtain $\mathbb{E}\Big[\frac{1}{\mu_0^\Phi} \Big]$ as follows:
\begin{align}
    &\mathbb{E}\Big[\frac{1}{\mu_0^\Phi} \Big] 
    \nonumber\\
    &\approx \int_0^\infty \!\! \exp \Big(\!-\!z\!
    \nonumber\\
    &\qquad +\!\frac{\lambda_\mathrm{s}}{\lambda_\mathrm{d}} \frac{\eta}{F} \Big(1\!\!-\!\! \frac{\eta}{F} \Big)^{\delta-1} \theta^\delta \Gamma(1\!-\!\delta) \Gamma(1\!+\!\delta) \Gamma(1\!+\!\epsilon) z^{1\!-\!\epsilon}\! \Big)\! \mathrm{d}z
    \nonumber\\
    &= \sum_{n=0}^\infty \frac{{C_2}^n}{n!}\Gamma \big(1+(1-\epsilon)n \big).
\end{align}
Substituting expressions for $\mathbb{E}[\mu_0^\Phi]$ and $\mathbb{E}\Big[\frac{1}{\mu_0^\Phi} \Big]$ into (50), we can get the approximate average AoI.

\end{appendix}
\bibliographystyle{IEEEtran}
\bibliography{bib/StringDefinitions,bib/IEEEabrv,bib/howard_AoI_Ctrl}

\end{document}